\documentclass[twocolumn,showpacs,superscriptaddress]{revtex4}

\usepackage[dvipdfmx]{graphicx}
\usepackage{amsmath}%
\usepackage{dcolumn}
\usepackage{bm}
\usepackage {cancel}
\usepackage{here}
\usepackage{epstopdf}
\usepackage{xcolor}

\let\Re\relax
\DeclareMathOperator{\Re}{Re}

\begin{document}

\title{Resonance fluorescence spectra of a driven Kerr nonlinear resonator}

\author{Aree~Taguchi}
\affiliation{
Division of Physics, Faculty of Pure and Applied Sciences, University of Tsukuba, Tennodai, Tsukuba, Ibaraki 305-8571, Japan
}
\affiliation{
NEC-AIST Quantum Technology Cooperative Research Laboratory, Tsukuba, Ibaraki 305-8568, Japan
}


\author{Katsuta~Sakai}
\altaffiliation[Present address: ]
{Department of Physics, Gakushuin University, 1-5-1 Mejiro, Toshima, Tokyo 171-8588, Japan, and KEK Theory Center, Institute of Particle and Nuclear Studies, 1-1 Oho, Tsukuba, Ibaraki 305-0801, Japan}
\affiliation{
Institute for Liberal Arts, Institute of Science Tokyo, Ichikawa, Chiba 272-0827, Japan
}

\author{Aiko~Yamaguchi}
\affiliation{
NEC-AIST Quantum Technology Cooperative Research Laboratory, Tsukuba, Ibaraki 305-8568, Japan
}
\affiliation{
NEC Secure System Platform Research Laboratories, Tsukuba, Ibaraki 305-8568, Japan
}

\author{Yuya~Kano}
\affiliation{
NEC-AIST Quantum Technology Cooperative Research Laboratory, Tsukuba, Ibaraki 305-8568, Japan
}
\affiliation{
NEC Secure System Platform Research Laboratories, Tsukuba, Ibaraki 305-8568, Japan
}

\author{Yohei~Kawakami}
\affiliation{
NEC-AIST Quantum Technology Cooperative Research Laboratory, Tsukuba, Ibaraki 305-8568, Japan
}
\affiliation{
NEC Secure System Platform Research Laboratories, Tsukuba, Ibaraki 305-8568, Japan
}

\author{Tomohiro~Yamaji}
\affiliation{
NEC-AIST Quantum Technology Cooperative Research Laboratory, Tsukuba, Ibaraki 305-8568, Japan
}
\affiliation{
NEC Secure System Platform Research Laboratories, Tsukuba, Ibaraki 305-8568, Japan
}
\affiliation{
RIKEN Center for Quantum Computing (RQC), Wako, Saitama 351-0198, Japan
}

\author{Tetsuro~Satoh}
\affiliation{
NEC-AIST Quantum Technology Cooperative Research Laboratory, Tsukuba, Ibaraki 305-8568, Japan
}
\affiliation{
NEC Secure System Platform Research Laboratories, Tsukuba, Ibaraki 305-8568, Japan
}
\affiliation{
RIKEN Center for Quantum Computing (RQC), Wako, Saitama 351-0198, Japan
}
\author{Ayuka~Morioka}
\affiliation{
NEC-AIST Quantum Technology Cooperative Research Laboratory, Tsukuba, Ibaraki 305-8568, Japan
}
\affiliation{
NEC Secure System Platform Research Laboratories, Tsukuba, Ibaraki 305-8568, Japan
}

\author{Kiyotaka~Endou}
\affiliation{
NEC-AIST Quantum Technology Cooperative Research Laboratory, Tsukuba, Ibaraki 305-8568, Japan
}
\affiliation{
NEC Secure System Platform Research Laboratories, Tsukuba, Ibaraki 305-8568, Japan
}

\author{Yuichi~Igarashi}
\affiliation{
NEC-AIST Quantum Technology Cooperative Research Laboratory, Tsukuba, Ibaraki 305-8568, Japan
}
\affiliation{
NEC Secure System Platform Research Laboratories, Tsukuba, Ibaraki 305-8568, Japan
}

\author{Masayuki~Shirane}
\affiliation{
NEC-AIST Quantum Technology Cooperative Research Laboratory, Tsukuba, Ibaraki 305-8568, Japan
}
\affiliation{
NEC Secure System Platform Research Laboratories, Tsukuba, Ibaraki 305-8568, Japan
}

\author{Yasunobu~Nakamura}
\affiliation{
RIKEN Center for Quantum Computing (RQC), Wako, Saitama 351-0198, Japan
}
\affiliation{
Department of Applied Physics, Graduate School of Engineering, The University of Tokyo, Bunkyo-ku, Tokyo 113-8656, Japan
}

\author{Kazuki~Koshino}
\affiliation{
Institute for Liberal Arts, Institute of Science Tokyo, Ichikawa, Chiba 272-0827, Japan
}

\author{Tsuyoshi~Yamamoto}
\affiliation{
Division of Physics, Faculty of Pure and Applied Sciences, University of Tsukuba, Tennodai, Tsukuba, Ibaraki 305-8571, Japan
}
\affiliation{
NEC-AIST Quantum Technology Cooperative Research Laboratory, Tsukuba, Ibaraki 305-8568, Japan
}
\affiliation{
NEC Secure System Platform Research Laboratories, Tsukuba, Ibaraki 305-8568, Japan
}
\affiliation{
RIKEN Center for Quantum Computing (RQC), Wako, Saitama 351-0198, Japan
}

\date{\today}

\begin{abstract}
Resonance fluorescence spectra of a driven Kerr nonlinear resonator is investigated both theoretically and experimentally.  
When the Kerr nonlinear resonator is driven strongly such that the induced 
Rabi frequency is comparable to or larger than the Kerr nonlinearity, 
the system cannot be approximated as a two-level system. We theoretically derive  
characteristic features in the fluorescence spectra such as the decrease of the center-peak intensity 
and the asymmetric sideband peaks in the presence of finite dephasing. 
Those features are consistently explained by the population of the initial dressed state and its 
transition matrix element to the final dressed state of the transition corresponding to each peak. 
Finally, we experimentally measure the resonance fluorescence spectra of a driven superconducting Kerr nonlinear resonator 
and find a quantitative agreement with our theory. 
\end{abstract}


\maketitle
\section{Introduction}
Resonance fluorescence is a fundamental phenomenon of quantum optics manifesting 
coherent light--matter interactions~\cite{Mollow_1969,Kimble_1976}. 
It was first observed in atomic systems~\cite{Schuda_1974,Wu_1975}, 
and more recently in artificial-atom systems such as semiconductor quantum dots~\cite{Muller_2007} 
and superconducting circuits~\cite{Astafiev_2010}. 

In superconducting systems, referred to as a circuit-quantum-electrodynamics system~\cite{Blais_2004,Gu_2017}, 
resonance fluorescence has been extensively studied in the contexts of photon blockade~\cite{Lang_2011}, 
creation of super- and subradiant states~\cite{vanLoo_2013}, qubit readout using squeezed microwaves~\cite{Toyli_2016}, 
and characterization of qubit decoherence rates~\cite{Lu_2021}. 
Most of these studies successfully treated the qubit as a two-level system with an exception of Ref.~\cite{Gasparinetti_2017}, 
where they observed two-photon resonance fluorescence involving a second excited state of a transmon qubit. 
However, superconducting qubits are essentially a multi-level system, and 
it is expected that such a two-level approximation becomes invalid if the qubit is driven so strongly 
that the induced Rabi frequency is comparable to the anharmonicity of the qubit.  
It is also worth mentioning that the resonance fluorescence in three-level systems with  
$\Lambda$-, $V$- and $\Xi$-type energy-level configurations 
has been studied both in theory~\cite{Fu_1992} and experiment in the atomic system~\cite{Tian_2012}. 

Energy levels of the dressed states in the strong-drive regime, 
where the Rabi frequency is comparable to the anharmonicity, 
has been probed spectroscopically in a transmon qubit~\cite{Bauer_2009,Koshino_2013} and a Kerr parametric oscillator (KPO)~\cite{Yamaji_2022}. 
However, fluorescence under the strong-drive regime has not been reported. 

In the present work, we theoretically and experimentally study the fluorescence from a nonlinear resonator 
with Kerr nonlinearity. 
In the theory, we develop a general formalism for the fluorescence spectrum, 
and apply it to a transmon qubit~\cite{Koch_2007} and a Kerr nonlinear resonator (KNR)~\cite{Goto_2016,Puri_2017}, 
which have different magnitudes of the Kerr nonlinearity, to compare them with an ideal two-level system. 
In the experiment, we use a KNR with the nonlinearity of around 10~MHz, which is one order smaller than 
that of a typical transmon qubit and is therefore suitable for realizing the strong-drive regime. 

The present paper is organized as follows. 
In Sec.~\ref{Sec_Theory}, we develop a general formalism of the coherent and incoherent scatterings of an external drive field by 
a KNR, latter of which gives fluorescence spectrum of the KNR. 
Using the theory, we calculate the resonance-fluorescence spectrum of a strongly driven KNR.  
We observe that the spectrum exhibits characteristic features, which are distinct from the cases of a two-level system and a 
weakly driven KNR. We also show that those features can be explained in terms of the state population 
and the transition matrix element for the corresponding transition. 
Section~\ref{Sec_Exp} describes our experiments. 
We measure the resonance fluorescence of a superconducting KNR 
and confirm that the observed spectrum is well reproduced by our theory. 
Finally, we conclude our discussion in Sec.~\ref{Sec_Con}. 

\section{Theory}~\label{Sec_Theory}
\subsection{Calculation of the fluorescence spectra}
\begin{figure}[t]
\includegraphics[width=0.9\columnwidth,clip]{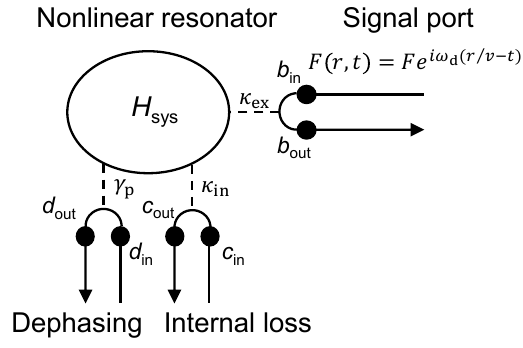}
\caption{~\label{fig1}
System under consideration. 
}
\end{figure}

Figure~\ref{fig1} shows a schematic diagram of the system we consider in this study. 
A Kerr nonlinear resonator (KNR) is coupled to the signal port, which is a semi-infinite transmission line 
to apply a drive field to the KNR. 
The KNR is also coupled to fictitious loss ports to account for its internal loss and dephasing. 
Here we assume dephasing due to the Markovian noise. 
The Hamiltonian of the system is given by 
\begin{eqnarray} \label{hamiltonian}
\mathcal{H} &=& \mathcal{H}_{\rm sys} + \mathcal{H}_{\rm sig} + \mathcal{H}_{\rm loss} + \mathcal{H}_{\rm dep}, \\ \label{H_sys}
\mathcal{H}_{\rm sys}/\hbar &=& \Omega_0 a^\dagger a + \frac{K}{2}a^\dagger a^\dagger aa, \\
\mathcal{H}_{\rm sig}/\hbar &=& \int dk \Bigl[v_b k b_k^\dagger b_k + 
\sqrt{\frac{v_b \kappa_{\rm ex}}{2 \pi}}\,
\Bigl( a^\dagger b_k + b_k^\dagger a \Bigr) \Bigr], \\
\mathcal{H}_{\rm loss}/\hbar &=& \int dk \Bigl[v_c k c_k^\dagger c_k + 
\sqrt{\frac{v_c \kappa_{\rm in}}{2 \pi}}\,
\Bigl( a^\dagger c_k + c_k^\dagger a \Bigr) \Bigr], \\
\mathcal{H}_{\rm dep}/\hbar &=& \int dk \Bigl[v_d k d_k^\dagger d_k + 
\sqrt{\frac{v_d \gamma_{\rm p}}{\pi}}\,
a^\dagger a \Bigl(d_k + d_k^\dagger \Bigr) \Bigr] \label{d_port}.  
\end{eqnarray}
Here, $\Omega_0$ and $K$ represent the resonance frequency and the Kerr nonlinearity of the resonator, respectively, 
and $a$ is an annihilation operator for the resonator mode, which satisfies $[a, a^\dagger]=1$. 
The field operator in the wave-number representation, $b_k$, is for the mode in the signal port 
with the wave number $k$ and velocity $v_b$. 
The coupling between the resonator and the signal port is denoted by $\kappa_{\rm ex}$, the external coupling rate. 
Similarly, $c_k$ and $d_k$ are field operators for the fictitious loss ports describing the internal loss and dephasing, respectively. 
Their coupling strengths, $\kappa_{\rm in}$ and $\gamma_{\rm p}$, correspond to the internal loss and dephasing rates, respectively. 
The operators $b_k$, $c_k$ and $d_k$ satisfy the following commutation rules: 
$\bigl[ b_k,b^\dagger_{k'} \bigr] =\bigl[ c_k,c^\dagger_{k'} \bigr] = \bigl[ d_k,d^\dagger_{k'} \bigr] = \delta(k-k')$. 
Hereafter, we assume that $v_b=v_c=v_d$ for simplicity and denote them by $v$. 

Because of the continuous driving, the KNR is expected to show fluorescence, which is contained in the $b_{\rm out}$ field as an incoherent component. 
In fact, an analytical formula of the fluorescence spectra for a two-level sysmtem
has been derived in Refs.~\cite{Koshino_2012, Lu_2021}. 
Here, we extend it for a multi-level system, namely, the Kerr nonlinear resonator. 
Below we show the outline of the calculation. The details are explained in Appendix~\ref{App1}. 
Note that, in contrast to the case of a two-level system, 
we cannot obtain a compact analytic expression for the fluorescence spectra for a KNR. 
They are calculated after solving a set of simultaneous equations involving an infinite number of variables by truncation. 

We assume that, at $t$=0, the overall system is in its ground state and 
a drive field $\mathcal{F}(r/v-t)$ is injected from the signal port, where
\begin{equation}
\mathcal{F}(r/v-t) = Fe^{\mathrm{i}\omega_{\rm d}(r/v-t)}/\sqrt{v}.
\end{equation}
Here, $\omega_{\rm d}$ is the angular frequency of the drive field and $F$ is its amplitude in units of $1/\sqrt{\mathrm s}$. 
Note that $|F|^2$ represents the photon rate. 
In the steady state, the power spectrum of the output field $b_{\rm out}(t)$ is given by
\begin{equation}
S(\omega) =
\Re \int_0^\infty \frac{d\tau}{\pi} e^{\mathrm{i}\omega\tau}
\langle b_{\rm out}^\dagger(t) b_{\rm out}(t+\tau)\rangle,  
\end{equation}
where the bracket represents the expectation value calulated with the initial state vector [Eq.~(\ref{isv})], 
which is the eigenstate of the input field $b_{\rm in}(t)$. 
Using the input--output relation 
\begin{equation}~\label{b_iot_m}
b_{\rm out}(t) = b_{\rm in}(t) - \mathrm{i} \sqrt{\frac{\kappa_{\rm ex}}{v}}\, a(t), 
\end{equation}
we obtain the following formula
\begin{eqnarray}
S(\omega) &\equiv& S_{\rm c}(\omega) + S_{\rm i}(\omega) \\ \label{Scoherent_m}
S_{\rm c}(\omega) &=& 
\Re \int_0^\infty \frac{d\tau}{\pi} e^{\mathrm{i}\omega\tau}
\langle b_{\rm out}^\dagger(t)\rangle
\langle b_{\rm out}(t+\tau)\rangle \\ \label{Sincoherent_m}
S_{\rm i}(\omega) &=&
\Re \int_0^\infty \frac{d\tau}{\pi} e^{\mathrm{i}\omega\tau}\frac{\kappa_{\rm ex}}{v}
\langle a^\dagger(t),a(t+\tau)\rangle, 
\end{eqnarray}
where $S_{\rm c}(\omega)$ and $S_{\rm i}(\omega)$ represent 
coherent and incoherent components, respectively, 
and $\langle A, B\rangle \equiv \langle AB\rangle - \langle A\rangle\langle B\rangle$. 

\subsubsection{Calculation of the two-time correlation function}
In order to evaluate $\langle a^\dagger(t),a(t+\tau)\rangle$ in $S_{\rm i}(\omega)$, 
we consider its equation of motion derivable from the Hamiltonian Eq.~(\ref{hamiltonian}). 
In contrast to the case of a two-level system [Eqs.~(\ref{B1_deriv2})--(\ref{B3_deriv})], 
the simultaneous equations of motion are not in a closed form. 
Therefore, we define the following quantity 
\begin{equation}~\label{B_mn_def}
B_{m,n}(t,\tau) \equiv \langle a^\dagger(t), A_{m,n}(t+\tau) \rangle, 
\end{equation}
where $A_{m,n} \equiv {a^\dagger}^m a^n$, and solve a set of simultaneous linear equations numerically 
with a truncation for $m$ and $n$.
Assuming a steady state, we further set $B_{m,n}$ as 
\begin{equation}~\label{B_mn_to_beta_mn}
B_{m,n}(t,\tau) = e^{\mathrm{i}(m-n)\omega_{\rm d}\tau}e^{\mathrm{i}(m-n+1)\omega_{\rm d}t} 
\beta_{m,n}(\tau).
\end{equation}
The set of simultaneous equations for $\overline{\beta_{m,n}}(s)$, which is a Laplace transform of $\beta_{m,n}(\tau)$, is given by 
\begin{widetext}
\begin{equation}~\label{L_eom_beta_m}
\beta_{m,n}(0) =
(s-\epsilon_{m,n}')\, \overline{\beta_{m,n}}(s)
-\mathrm{i}K(m-n)\, \overline{\beta_{m+1,n+1}}(s)
 + \mathrm{i}\sqrt{\kappa_{\rm ex}}\, [n\, \overline{\beta_{m,n-1}}(s) F-
m\, \overline{\beta_{m-1,n}}(s)F^*]. 
\end{equation}
\end{widetext}
\subsubsection{Calculation of the one-time correlation function}
In order to solve the simultaneous equations represented by Eq.~(\ref{L_eom_beta_m}), 
we first need to determine the left-hand side, $\beta_{m,n}(0)$. 
From Eqs.~(\ref{B_mn_def}) and (\ref{B_mn_to_beta_mn}), 
we have $\beta_{m,n}(0)=\Bigl[ \langle A_{m+1,n}(t) \rangle -\langle A_{1,0}(t) \rangle\langle A_{m,n}(t) \rangle \Bigr]e^{-\mathrm{i}(m-n+1)\omega_{\rm d}t}$. 
Therefore, $\beta_{mn}(0)$ is determined from the set of $\langle A_{m,n}(t) \rangle$, the one-time correlation functions.
They are obained by solving a set of simultaneous linear equations of motion for $A_{m,n}(t)$ in the steady state. 
Setting $\langle A_{m,n}(t)\rangle$ as 
\begin{equation}~\label{alpha_mn_s}
\langle A_{m,n}(t)\rangle = \langle\alpha_{m,n}\rangle_{\rm s} e^{\mathrm{i}(m-n)\omega_{\rm d}t}, 
\end{equation}
the set of simultaneous equations for $\langle\alpha_{m,n}\rangle_{\rm s}$ is given by 
\begin{widetext}
\begin{equation}
[\epsilon_{m,n} - \mathrm{i}(m-n)\omega_{\rm d}] \langle \alpha_{m,n}\rangle_{\rm s} +\mathrm{i}K(m-n) \langle \alpha_{m+1,n+1}\rangle_{\rm s}
- \mathrm{i}\sqrt{\kappa_{\rm ex}}\, [n \langle \alpha_{m,n-1}\rangle_{\rm s} F
- m \langle \alpha_{m-1,n}\rangle_{\rm s} F^*] =0,
\end{equation}
\end{widetext} 
where
$\epsilon_{m,n} \equiv 
\mathrm{i}(m-n)\Omega_0 + \mathrm{i}(m-n)(m+n-1)\frac{K}{2} 
-(m+n)\frac{\kappa_{\rm ex}+\kappa_{\rm in}}{2} - (m-n)^2\gamma_{\rm p}.$ 
From these equations, $\langle \alpha_{m,n}\rangle_{\rm s}$ is numerically obtained.
By noting that $\beta_{m,n}(0) = \langle \alpha_{m+1,n} \rangle_{\rm s} - \langle \alpha_{1,0} \rangle_{\rm s}\langle \alpha_{m,n} \rangle_{\rm s}$ in Eq.~(\ref{L_eom_beta_m}), 
$\overline{\beta_{m,n}}(s)$ is obtained by solving 
\begin{widetext}
\begin{multline}~\label{L_eom_beta_alpha_m}
\langle \alpha_{m+1,n} \rangle_{\rm s} - \langle \alpha_{1,0} \rangle_{\rm s} \langle \alpha_{m,n} \rangle_{\rm s}=
(s-\epsilon’_{m,n})\, \overline{\beta_{m,n}}(s)
-\mathrm{i}K(m-n)\, \overline{\beta_{m+1,n+1}}(s)\\
 + \mathrm{i}\sqrt{\kappa_{\rm ex}}\, [n\, \overline{\beta_{m,n-1}}(s) F- m\, \overline{\beta_{m-1,n}}(s)F^*] 
\end{multline}
\end{widetext}
for $m,n=0,1,\cdots, N_{\rm max}$, 
where $\epsilon_{m,n}' \equiv \epsilon_{m,n} - \mathrm{i}(m-n)\omega_{\rm d}$ 
and we typically set $N_{\rm max}=20$. 

\subsubsection{Calculation of the output-field spectrum}
The incoherent component of the output-field spectrum is obtained 
from Eq.~(\ref{Sincoherent_m}) as 
\begin{equation}~\label{Sincoherent_m2}
S_{\rm i}(\omega)
= \frac{\kappa_{\rm ex}}{\pi v}\Re \overline{\beta_{0,1}}(\mathrm{i}(\omega_{\rm d}-\omega)). 
\end{equation}
We use the solution of $\overline{\beta_{0,1}}(s)$ in Eq.~(\ref{L_eom_beta_alpha_m}) and plug it into the above equation to 
obtain the power spectrum of the incoherent component, namely, the fluorescence. 

As for the coherent component, 
by using $\langle b_{\rm in}(t)\rangle = Fe^{-\mathrm{i}\omega_{\rm d}t}/\sqrt{v}$ and Eq.~(\ref{b_iot_m}), we obtain
\begin{equation}
\langle b_{\rm out}^\dagger(t)\rangle
\langle b_{\rm out}(t+\tau)\rangle
= \frac{e^{-\mathrm{i}\omega_{\rm d}\tau}}{v} \Bigl|F - \mathrm{i} \sqrt{\kappa_{\rm ex}}\, \langle \alpha_{0,1}\rangle_{\rm s}\Bigr|^2.
\end{equation}
Substituting this into Eq.~(\ref{Scoherent_m}), the spectrum is calculated to be 
\begin{equation}~\label{Scoherent_m2}
S_{\rm c}(\omega)
= \frac{\delta(\omega-\omega_{\rm d})}{v}\Bigl|F - \mathrm{i} \sqrt{\kappa_{\rm ex}}\, \langle \alpha_{0,1}\rangle_{\rm s}\Bigr|^2. 
\end{equation}

\begin{figure*}[t]
\includegraphics[width=1.8\columnwidth,clip]{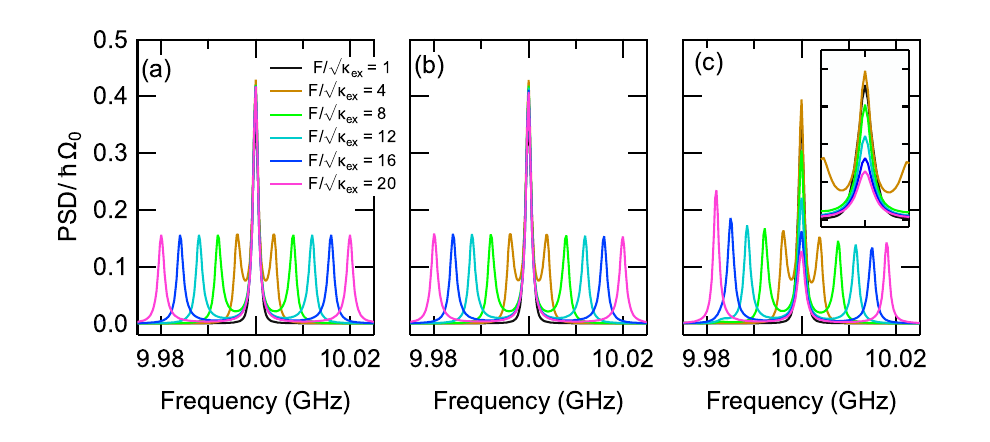}
\caption{~\label{fig2}
Spectra of resonance fluorescence for (a) a two-level system, (b) a nonlinear resonator 
in the transmon regime ($K/2\pi=-200$~MHz) and (c) a nonlinear resonator in the KNR regime ($K/2\pi=-20$~MHz). 
Other parameters used in the calculation are $\Omega_0/2\pi=10$~GHz, 
$\kappa_{\rm ex}=\kappa_{\rm in}=2\pi \times 0.5$~MHz, and $\gamma_{\rm p}/2\pi= 0.1$~MHz. 
}
\end{figure*}

\subsection{Numerical results}
Figure~\ref{fig2} shows the numerical results of the resonance fluorescence spectra with varied drive amplitude $F$ 
for three systems:  (a) a two-level system (TLS), (b) a nonlinear resonator with $K/2\pi=-200$~MHz (transmon)
and (c) a nonlinear resonator with $K/2\pi=-20$~MHz~(KNR). 
We assumed the dephasing rate of $\gamma_{\rm p}/2\pi$=0.1~MHz besides the external and internal loss rates of 
$\kappa_{\rm ex, in}/2\pi$=0.5~MHz. 
The spectra for the TLS and transmon [Figs.~\ref{fig2}(a) and \ref{fig2}(b)] look almost identical 
and exhibit a typical Mollow triplet consisting of  a center peak and symmetric sideband peaks.
As the theory predicts~\cite{Mollow_1969}, when $|F|^2$ is much larger than decay rates,  
the height of these peaks are almost independent of $F$, 
and the position of the sideband peaks linearly depend on $F$. 
Since the Rabi frequency is much smaller than the Kerr nonlinearity, the transmon is in the weak-drive regime in the calculated
range of $F$ and can be well approximated as a two-level system.  

The spectra for the KNR [Fig.~\ref{fig2}(c)] looks different and exhibit several distinct features. 
First, the heights of the sideband peaks are not symmetric: the lower sideband is higher than the upper sideband. 
Note that here we consider the resonant drive, namely, $\omega_{\rm d}=\Omega_0$. 
Asymmetric side peaks in fluorescence spectrum have been reported 
both in theory~\cite{Koshino_2012} and experiment~\cite{Lu_2021}, but they are under an off-resonant drive. 
Second, as shown in the inset of Fig.~2(c), the height of the center peak decreases significantly as $F$ increases. 
Third, the position of the sideband peaks is slightly different from those of the TLS and transmon, 
especially at large $F$, indicating the nonlinear dependence of the Rabi frequency on the drive strength~\cite{Bauer_2009}. 
Since the Rabi frequency is comparable to the Kerr nonlinearity, KNR is in the strong-drive regime for this range of $F$. 

\begin{figure*}[t]
\includegraphics[width=1.8\columnwidth,clip]{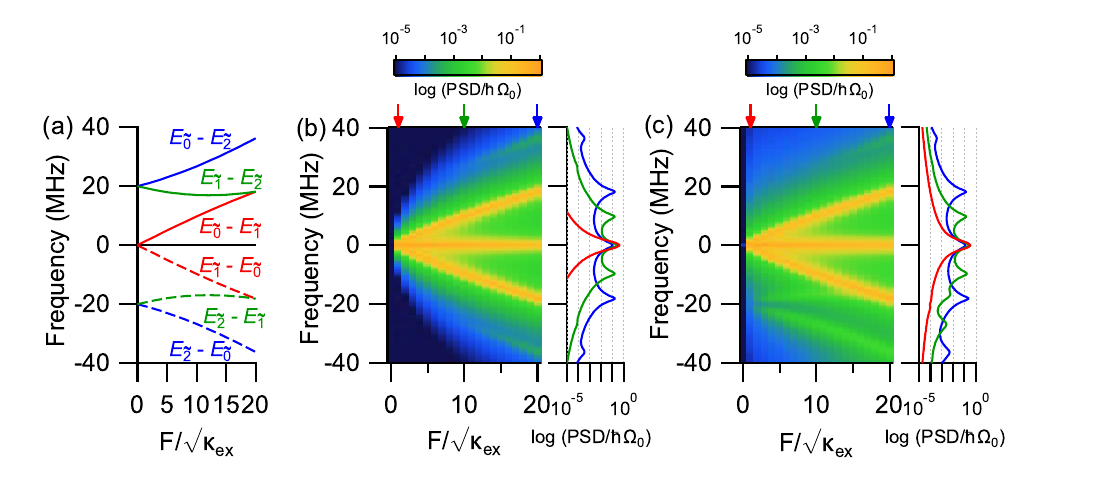}
\caption{~\label{fig3}
(a) Transition frequencies between the dressed states as a function of the drive amplitude $F$ normalized by 
$\sqrt{\kappa_{\rm ex}}$. 
(b) and (c) Resonance fluorescence spectra of the nonlinear resonator in the KNR regime 
as a function of the frequency and $F/\sqrt{\kappa_{\rm ex}}$. 
We set the dephasing rate $\gamma_{\rm p}/2\pi$=0 in (b) and 0.1~MHz in (c). 
Other parameters used in the calculation are the same as those in Fig.~\ref{fig2}(c). 
The frequency is measured from $\Omega_0/2\pi=10$~GHz. The right panels show the cross section of the left panels 
at the three values of $F/\sqrt{\kappa_{\rm ex}}$ indicated by the arrows. 
}
\end{figure*}

To further study the spectra of the KNR, we investigate the effect of dephasing. 
Figures~\ref{fig3}(b) and \ref{fig3}(c) show the resonance fluorescence spectra of a KNR calculated 
without and with the dephasing of $\gamma_{\rm p}/2\pi$=0.1~MHz, respectively. Other parameters used in the calculation 
are the same as those in Fig.~\ref{fig2}(c). 
As seen in Fig.~\ref{fig3}(b), the asymmetry in the spectrum disappears when there is no dephasing. 
We also plot in Fig.~\ref{fig3}(a) the transition frequencies between the dressed states as a function of $F$. 
The label in the figure $E_{\tilde{i}}-E_{\tilde{j}}$ represents the energy difference between $i$th and $j$th eigenstates, 
which are obtained by diagonalizing the following Hamiltonian under a frame rotating at $\omega_{\rm d}=\Omega_0$: 
\begin{equation}~\label{H_rot}
\mathcal{H}=\frac{K}{2}a^\dagger a^\dagger aa + \sqrt{\kappa_{\rm ex}}\, (F^*a + Fa^\dagger).
\end{equation}
By comparing the figure with the spectra, we see that the peaks in Figs.~\ref{fig3}(b) and \ref{fig3}(c) are observed 
at the transition frequencies between the dressed states. Especially, we observe more peaks than 
in the case of a TLS, namely, the Mollow triplet. 
They are conspicuous in the lower frequency range of Fig.~\ref{fig3}(c) and attributed to the transitions from 
the second excited state. 

\begin{figure*}[t]
\includegraphics[width=1.8\columnwidth,clip]{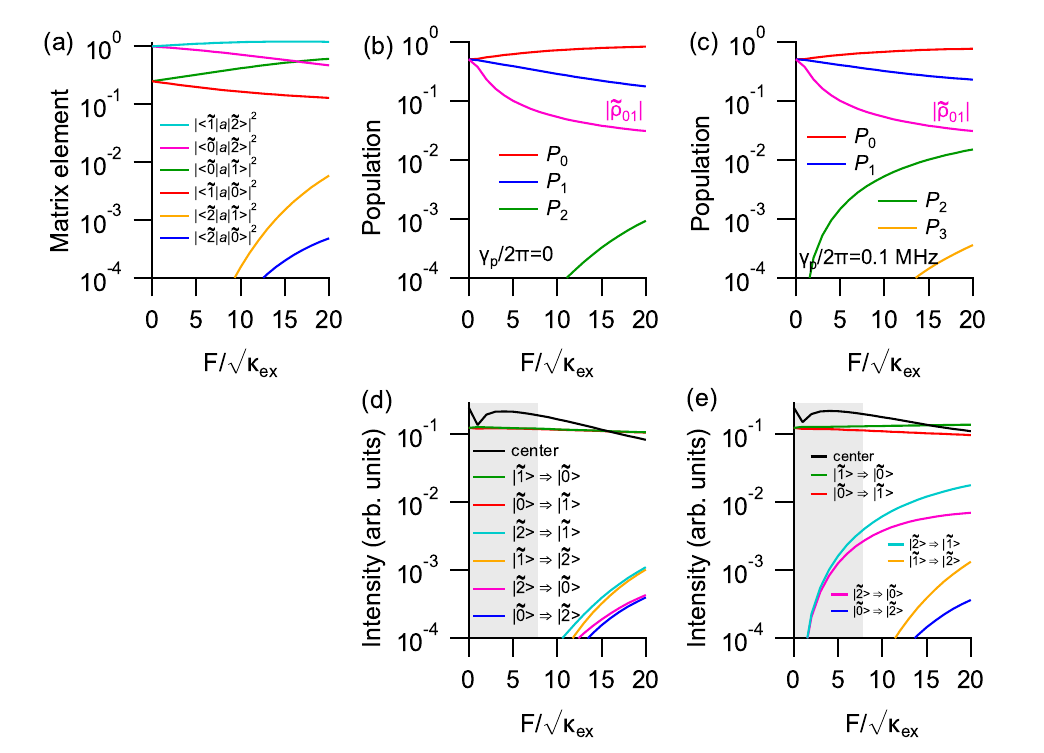}
\caption{~\label{fig4}
(a) Squared transition-matrix elements $|\langle \tilde{j}|a|\tilde{i}\rangle|^2$ between the dressed states of the KNR. 
Eigenvectors are calculated by diagonalizing the Hamiltonian in Eq.~(\ref{H_rot}). 
(b) and (c) Steady-state populations of the eigenstates, $P_i$, of the KNR for $\gamma_{\rm p}/2\pi=0$ and 0.1~MHz, respectively.
Other parameters used in the calculation are the same as those in Fig.~\ref{fig2}(c). 
Absolute value of the (0,1) component of the density matrix $|\tilde{\rho}_{01}|$ is also plotted. 
(d) and (e) Estimate of the fluorescence peak intensity from the product of $P_i$ and $|\langle \tilde{j}|a|\tilde{i}\rangle|^2$ 
for $\gamma_{\rm p}/2\pi=0$ and 0.1~MHz, respectively. 
The center-peak intensity (black line) is calculated with Eq.~(\ref{center_peak_int}).
The gray areas indicate the region
where the off-diagonal components of the density matrix are non-negligible 
[$|\tilde{\rho}_{01}|$ larger than 10\% of max($|\tilde{\rho}_{00}|$, $|\tilde{\rho}_{11}|$)] 
and therefore estimation based on Eq.~(\ref{Si_int}) becomes unreliable.
}
\end{figure*}

To obtain insight into these characteristics, we adopt the notion 
that the peak intensity in the fluorescence is given by the product of the population of the initial state 
and the transition-matrix element between the initial and final states of the corresponding transition~\cite{Lu_2021}. 
From Eq.~(\ref{Sincoherent}), the total spectrum intensity is given by
\begin{eqnarray}~\nonumber~\label{Si_int}
\int \! d\omega \, S_{\rm i}(\omega) &\propto& \langle a^\dagger a \rangle_{\rm s} - |\langle a\rangle_{\rm s}|^2\\
&\approx& \sum_{i,j} P_i  |\langle \tilde{j}|a|\tilde{i}\rangle|^2- |\langle a\rangle_{\rm s}|^2, 
\end{eqnarray}
where $P_i$ represents the population of the dressed state $|\tilde{i}\rangle$.
Note that $P_i$ is a diagonal element of the density matrix. 
The last expression in Eq.~(\ref{Si_int}) neglects contribution from the nondiagonal elements. 
Those are, however, 
negligibly smaller than the diagonal ones as we observe in Figs.~\ref{fig4}(b) and \ref{fig4}(c) 
in the regime of $|F|^2>\kappa_{\rm ex}$.

Figure~\ref{fig4}(a) shows the transition-matrix elements between the dressed states as a function of $F/\sqrt{\kappa_{\rm ex}}$. 
They are calculated using the eigenvectors of the Hamiltonian Eq.~(\ref{H_rot}). 
Figures~\ref{fig4}(b) and \ref{fig4}(c) show the population of the dressed states 
when $\gamma_{\rm p}/2\pi=0$ and 0.1~MHz, respectively. 
The population is calculated from $\langle\alpha_{m,n}\rangle_{\rm s}$ in Eq.~(\ref{alpha_mn_s}) using the relation in Appendix~\ref{App2}.  
As seen in the figures, dephasing enhances the population of higher levels. 
We also plot the absolute value of the (0,1) component of the density matrix $|\tilde{\rho}_{01}|$ calculated with Eq.~(\ref{alpha_s_3}). 
It is confirmed that they are much smaller than the dominant populations $P_0$ and $P_1$ 
when $F$ is sufficiently larger than $\sqrt{\kappa_{\rm ex}}$.
Figures~\ref{fig4}(d) and \ref{fig4}(e) show the product of the population $P_i$ and the transition matrix element
$|\langle \tilde{j}|a|\tilde{i}\rangle|^2$, which is expected to represent the intensity of the fluorescence peak corresponding to the 
transition from $|\tilde{i}\rangle$ to $|\tilde{j}\rangle$. 
Also plotted in Figs.~\ref{fig4}(c) and \ref{fig4}(d) in black lines are the intensity of the center peak, 
which is obtained by 
\begin{equation}~\label{center_peak_int}
\sum_iP_i |\langle \tilde{i}|a|\tilde{i}\rangle|^2 - |\langle a \rangle_{\rm s}|^2.
\end{equation}

As seen from Fig.~\ref{fig4}(d), the intensities of the paired sidebands of 
$|\tilde{1}\rangle \rightarrow |\tilde{0}\rangle$ and $|\tilde{0}\rangle \rightarrow |\tilde{1}\rangle$ are almost equal and 
independent of $F$ in this range, which is consistent with the result in Fig.~\ref{fig3}(b). 
From Figs.~\ref{fig4}(a) and \ref{fig4}(b), it is seen that this fact originates in the balance of 
the population and the transition matrix element. 
Namely, the increase in $P_0$ is compensated by the decrease in $|\langle \tilde{1}|a|\tilde{0}\rangle|^2$, and 
the decrease in $P_1$ is compensated by the increase in $|\langle \tilde{0}|a|\tilde{1}\rangle|^2$. 
It is also worth mentioning that intensities of other pairs of sideband 
(e.g., $|\tilde{2}\rangle \rightarrow |\tilde{1}\rangle$ and $|\tilde{1}\rangle \rightarrow |\tilde{2}\rangle$)
are almost equal and are thus symmetric. 

In the presence of dephasing ($\gamma_{\rm p}/2\pi=0.1$~MHz), 
we see in Fig.~\ref{fig4}(c) that the populations of the excited states increase. 
This breaks the balance realized in the case without dephasing. 
As seen in Fig.~\ref{fig4}(e), the intensity of the peak corresponding 
$|\tilde{1}\rangle \rightarrow |\tilde{0}\rangle$ ($|\tilde{0}\rangle \rightarrow |\tilde{1}\rangle$)
increases (decreases) as a function of $F$. 
Also, intensities of other pairs of sideband are not equal and  
the intensity of the lower sideband peak is higher than the higher sideband peak
(e.g., $|\tilde{2}\rangle \rightarrow |\tilde{1}\rangle$ is higer than $|\tilde{1}\rangle \rightarrow |\tilde{2}\rangle$). 
These are consistent with what is observed in Fig.~\ref{fig3}(c). 

More intuitively, the asymmetry of the sideband-peak intensity in the presence of dephasing 
can be understood in the following way. 
In general, dephasing in material quantum systems turns imaginary excitations to real excitations and 
accordingly makes their fluorescence spectra closer to the spontaneous-emission spectra~\cite{Koshino_2011}. 
In a KNR, because of the negative Kerr nonlinearity ($K<0$), 
the higher transition frequencies ($|j\rangle$ and $|j+1\rangle$ for $j\ge1$) 
are smaller than the lowest transition frequency $\Omega_0$ ($|0\rangle$ and $|1\rangle$), 
so the spontaneous-emission spectrum from a highly-excited state would 
have a peak at a frequency lower than $\Omega_0$. 
Therefore, in the present situation where a KNR is driven by a resonant drive field at $\Omega_0$, 
the lower sideband of the Mollow triplet is more enhanced than the higher one. 
This intuitive understanding is supported by our numerical simulation (data not shown), 
where the higher sideband becomes more enhanced 
than the lower one when we assume a positive Kerr nonlinearity ($K>0$). 

As for the center peak, the intensity decreases with $F/\sqrt{\kappa_{\rm ex}}\gtrsim5$ regardless of the dephasing. 
This is consistent with the results in Figs.~\ref{fig2}(c), \ref{fig3}(b) and \ref{fig3}(c), although it is less clear in the latter two. 
Intuitively, a KNR, due to its weak nonlinearity, becomes more like a harmonic oscillator at higher drive strength, 
which does not have incoherent scattering. The decrease in the intensity of the center peak is 
attributable to the energy transfer from the incoherent to coherent scatterings [Eqs.~(\ref{Sincoherent_m}) and (\ref{Scoherent_m})]. 
Note that the finite intensity including the increase of the center peak at $F\rightarrow 0$ in Figs.~\ref{fig4}(d) and \ref{fig4}(e) is 
due to the breakdown of  the approximation in Eq.~(\ref{Si_int}).
Ideally, all the peaks should vanish at $F\rightarrow 0$. 

\section{Experiments}~\label{Sec_Exp}
To compare the above theory with experiments, 
we measure the fluorescence spectra in the strong-drive regime 
using a KNR with the Kerr nonlinearity of about 10~MHz. 
The KNR device used in the experiment is a type of frequency-tunable Xmon~\cite{Barends_2013}
with the same design as the one reported in Ref.~\cite{Yamaguchi_2024}, 
which consists of niobium electrodes sputtered and patterned on a silicon substrate and shadow-evaporated 
Josephson junctions made of aluminum. 
The KNR is capacitively coupled to the open end of a transmission line. 
In the present paper, we fix the magnetic flux bias of the KNR at zero.

\begin{figure}[t]
\includegraphics[width=0.9\columnwidth,clip]{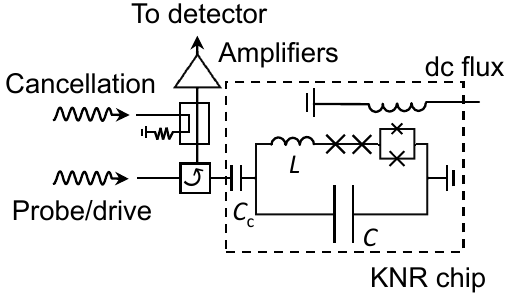}
\caption{~\label{fig5}
Simplified experimental setup. The KNR consists of an asymmetric SQUID in series with two Josephson junctions and 
a line with the geometrical inductance $L$, which are in parallel with a shunt capacitor with the capacitance $C$. 
The KNR is coupled to a semi-infinite transmission line via a coupling capacitor with the capacitance $C_{\rm c}$ 
in a reflection-type configuration, 
where the input and output fields are separated by a circulator. 
A cancellation filed is injected through a directional coupler. 
The amplifiers include an IMPA at 10 mK, a HEMT at 4 K and another at room temperature. 
}
\end{figure}

Figure~\ref{fig5} shows the simplified experimental setup for our measurements 
(See Appendix~\ref{App3} for the detailed one). 
The KNR is placed at the mixing-chamber stage of a dilution refrigerator at approximately 10~mK. 
For the measurement of reflection coefficients, we use a vector network analyzer. 
For the measurement of fluorescence spectra, 
%
a drive field is generated by a signal generator at room temperature and injected into the KNR. 
Fluorescence spectra from the KNR are measured using a spectrum analyzer with a resolution bandwidth of 10~kHz.
To improve the measurement signal-to-noise ratio (SNR), we repeatedly turn on and off the drive field to obtain their difference. 
Each scan is averaged for 10 times, and this process is repeated for 200 times. 
We also used an impedance-matched Josephson parametric amplifier (IMPA)~\cite{Urade_2021} 
for the detection of the fluorescence signal. 
The input 1-dB compression point of the IMPA is around $-110$~dBm. 
To avoid saturating IMPA by the reflected drive field, a cancellation field is injected through a directional coupler. 
Any reflection signal that could not be completely suppressed by this cancellation field was removed during the data analysis.

\begin{figure}[t]
\includegraphics[width=0.9\columnwidth,clip]{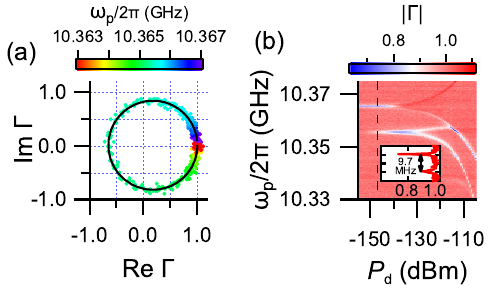}
\caption{~\label{fig6}
(a) Reflection coefficient $\Gamma$ of the KNR biased at zero magnetic field 
as a function of the probe frequency $\omega_{\rm p}$ plotted in an IQ plane. 
The power of the probe field is $-160.5$~dBm. 
The dots represent the experimental data, while the black line represents the fitting. 
(b) Two-tone spectroscopy, where $|\Gamma|$ is plotted as a function of $\omega_{\rm p}$ (vertical axis) 
under the application of an additional drive field at $\omega_{\rm d}=\Omega_0$ 
with varying power $P_{\rm d}$ (horizontal axis). 
The power of the probe field is $-140.5$~dBm.
The inset shows a cross-sectional view along the dotted line. 
The frequency separation between the two dips corresponds to the Kerr nonlinearity, which is $K/2\pi=-9.7$~MHz.
}
\end{figure}

Figure~\ref{fig6}(a) shows the frequency dependence of the reflection coefficient $\Gamma$ of the KNR biased at zero magnetic field, 
which is measured with a weak probe power $P_{\rm p}$ of $-160.5$~dBm.  
Throughout the paper, the powers of the various tones are specified at the sample chip, which is calibrated 
by the method described below. 
By fitting the data with 
\begin{equation}
\Gamma=1-\frac{\kappa_{\rm ex}}{\mathrm{i}(\omega_{\rm p}-\Omega_0)+\frac{\kappa_{\rm ex}+\kappa_{\rm in}^*}{2}}, 
\end{equation}
we obtain $\Omega_0/2\pi=10.3653$~GHz, $\kappa_{\rm ex}/2\pi=0.260 \pm 0.002$~MHz, 
and $\kappa_{\rm in}^*/2\pi=0.053 \pm 0.004$~MHz, after averaging 13 such measurements, 
where the errors are the standard deviations. 
Here,  $\kappa_{\rm in}^*$ represents the nominal internal loss rate obtained from the reflection coefficient, which 
is the sum of the actual internal loss rate and the twice of the pure dephasing rate $\gamma_{\rm p}$, i.e., 
$\kappa_{\rm in}^* = \kappa_{\rm in} + 2\gamma_{\rm p} $~\cite{Yamaguchi_2024}.

Figure~\ref{fig6}(b) shows the result of the two-tone spectroscopy, in which 
the reflection coefficient is measured under the application of an additional drive field at $\omega_{\rm d}=\Omega_0$ 
with varying powers $P_{\rm d}$s. 
This measurement reveals the energy spectrum of the dressed states formed by the KNR and the drive field~\cite{Yamaji_2022}. 
At weak drive powers, we observe two $P_{\rm d}$-independent dips one of which is located at $\omega_{\rm p}=\Omega_0$. 
The other one corresponds to the $|1\rangle$ to $|2\rangle$ transition of the bare KNR states
and their separation gives the Kerr nonlinearity, 
which is determined to be $-9.7$~MHz as shown in the inset. 

\begin{figure}[t]
\includegraphics[width=0.9\columnwidth,clip]{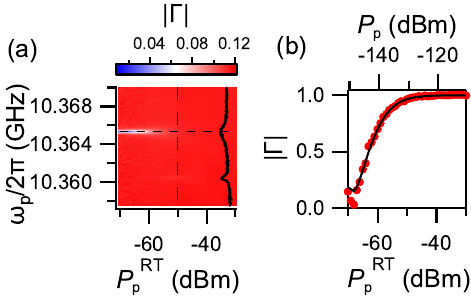}
\caption{~\label{fig7}
(a) Probe-power dependence of the reflection coefficient. 
$|\Gamma|$ is plotted as a function of the probe frequency $\omega_{\rm p}$ (vertical axis) and 
the probe power at room temperature $P_{\rm p}^{\rm RT}$ (horizontal axis). 
The horizontal dashed line represents $\omega_{\rm p}=\Omega_0$. 
The black curve represents the cross-section along the vertical dashed line, 
where the horizontal amplitude shows $|\Gamma|$ in a linearly ascending unit. 
(b) Probe-power dependence of $|\Gamma|$ at $\omega_{\rm p}=\Omega_0$. 
The red dots represent the experimental data, 
which is the normalized cross-section along the horizontal dashed line in~(a) plotted in the bottom axis, 
while the black line represents the theoretical curve~\cite{Yamaji_2022} plotted as a function of $P_{\rm p}$ in the top axis. 
In the calculation, we used $\Omega_0$, $\kappa_{\rm ex}$ and $\kappa_{\rm in}^*$ obtained from the measurement of $\Gamma$, 
and $K$ obtained from two-tone spectroscopy.
From the difference between $P_{\rm p}^{\rm RT}$ and $P_{\rm p}$, 
we determine the attenuation of $-80.5$~dB in the input line.  
}
\end{figure}

In order to determine the power of the fluorescence at the chip, we need to precisely determine the total amplification 
in the output line from the KNR chip to the detector at the room temperature. 
For that purpose, we calibrate the attenuation in the input line from the room temperature generator to the KNR chip 
by measuring the probe-power dependence of $\Gamma$ [Fig.~\ref{fig7}(a)]. 
Because we know the total attenuation of the measurement line consisting of input and output lines from $|\Gamma|$ off the 
resonance frequency [background amplitude of $|\Gamma|$ in Fig.~\ref{fig7}(a)], 
they allow us to determine the amplification in the output line. 

Figure~\ref{fig7}(a) shows $|\Gamma|$ as a function of $\omega_{\rm p}$ (vertical axis) and $P_{\rm p}^{\rm RT}$ (horizontal axis), 
where $P_{\rm p}^{\rm RT}$ is the power of the probe field represented by the value at the output of the network analyzer. 
As we increase $P_{\rm p}^{\rm RT}$, the dip observed at $\omega_{\rm p}=\Omega_0$ becomes shallower, and 
another dip at $\omega_{\rm p}=\Omega_0 - K/2$ corresponding to the two-photon transition from $|0\rangle$ to $|2\rangle$ states 
appears at around $P_{\rm p}^{\rm RT}=-50$~dB (vertical dashed line). 
By comparing $|\Gamma|$ at $\omega_{\rm p}=\Omega_0$ with the theory based on the input--output formalism~\cite{Yamaji_2022} 
using experimentally determined parameters of $K$, $\kappa_{\rm ex}$ and $\kappa_{\rm in}$ as shown in Fig.~\ref{fig7}(b), 
we determine the attenuation in the input line to be $-80.5$~dB. 
Note that in the calculation, we set $\kappa_{\rm in}=\kappa_{\rm in}^*$, namely, $\gamma_{\rm p}=0$. 
We confirmed that the result hardly depends on $\gamma_{\rm p}$ at least in a range relevant to the present study. 
From the total attenuation of $-18.9$~dB and 
taking into account the small difference in the attenuation between the setups 
for the measurement using network analyzer and spectrum analyzer, 
the gain in the output line is determined to be 61.0~dB. We also obtained consistent value in another calibration method 
based on the $P_{\rm d}$-dependence of the sideband-peak frequencies in the fluorescence spectra. 
Using this gain and that of the IMPA measured independently, we determine the amplitude of the measured spectra shown below. 

\begin{figure*}[t]
\includegraphics[width=1.8\columnwidth,clip]{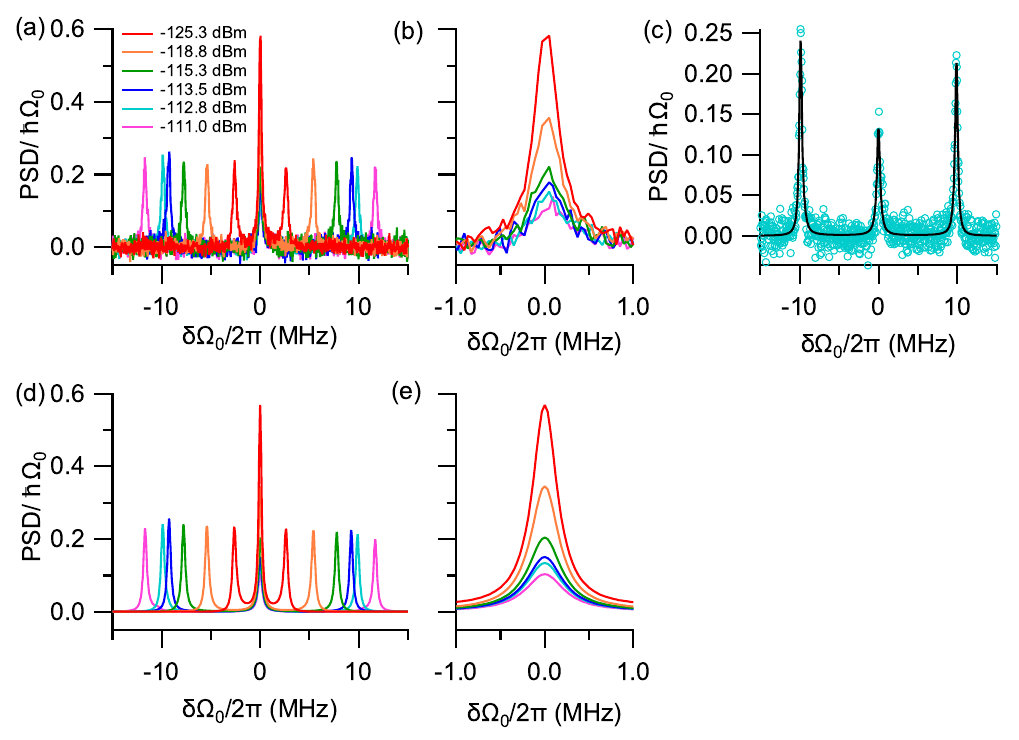}
\caption{~\label{fig8}
(a) Resonance fluorescence spectra of a KNR at various drive powers. 
The resonance frequency of the KNR is 10.365~GHz. 
(b) Magnified view of (a) around the center peak. 
(c)~Resonance fluorescence spectrum at $P_{\rm d}=-112.8$~dBm.
The circles represent the experimental data, and the black line represents a theoretical fit to the data. 
(d) Calculated spectra using the parameters
obtained from the fitting in (c) at the drive powers corresponding to the data in (a). 
(e) Magnified view of (d) around the center peak. 
}
\end{figure*}

Figure~\ref{fig8} shows the fluorescence spectra measured at different $P_{\rm d}$'s. 
Similar to the case of two-level systems, we observe one center peak at $\Omega_0$ and two sideband peaks, 
whose separation (Rabi frequency) increases with $P_{\rm d}$. 
Additional subpeaks due to higer energy states were not observed probably due the limited SNR.  
Particularly noteworthy is the clear decrease of the center-peak intensity with $P_{\rm d}$ as shown in Fig.~\ref{fig8}(b). 
This is one of the characteristic features of the strong-drive regime as we saw in Section~\ref{Sec_Theory}.  
Although the asymmetry of the sideband-peak intensities is another characteristic feature of the strong-drive regime, 
it is not so pronounced because the KNR is biased at the zero flux, where the pure dephasing rate is minimal. 
We pick up the data at $P_{\rm d}=-18.0$~dBm, where the Rabi frequency is close to $K$ and the large asymmetry 
in the sideband-peak intensities is expected in theory [Fig.~\ref{fig2}(c)], and 
fitted it with the theory [Eq.~(\ref{Sincoherent_m2})] as shown in Fig.~\ref{fig8}(c). 
In the fitting, we used $\Omega_0$, $K$, $\kappa_{\rm ex}$ and the drive strength $F$ as fixed parameters, 
which are determined from 
the experiments and the calibration, and we used 
the internal loss rate $\kappa_{\rm in}$, and the pure dephasing rate $\gamma_{\rm p}$ 
as fitting parameters. In addition, we introduced another fitting parameter $A$, the overall scaling factor to account for the 
small error in the calibration of the amplification factor in the output line. 
We used the python package Optuna~\cite{Akiba_2019} for the fitting. 
We set the range of fitting parameters as $0\le\kappa_{\rm in}\le2\pi\times0.07$~MHz, 
$0\le\gamma_{\rm p}\le2\pi\times0.01$~MHz and $0.7\le A\le1.1$ and the number of evaluation trials as 500.
As a result of the fitting, we obtained $\kappa_{\rm in}/2\pi=0.046$~MHz, $\gamma_{\rm p}/2\pi=0.0065$~MHz, and $A=0.83~(-0.8$~dB). 
Note that $\kappa_{\rm in} + 2\gamma_{\rm p}=2\pi \times 0.059$~MHz, which is consistent with $\kappa_{\rm in}^*$ 
obtained from the measurement of $\Gamma$. 
 
Using the parameters obtained from the fitting, 
we calculate the spectra based on Eq.~(\ref{Sincoherent_m2}) at different drive powers and plot them in Figs.~\ref{fig8}(d) and (e). 
The overall behavior of the $P_{\rm d}$-dependence of the sideband peaks and the center peak 
agrees well with the experimental results in Figs.~\ref{fig8}(a) and (b). 

\section{Conclusion}~\label{Sec_Con}
We have investigated the fluorescence spectrum of the driven KNR. 
First, we developed a general formalism of coherent and incoherent scatterings of an external drive field by a KNR. 
The incoherent scattering gives the fluorescence spectrum of the driven KNR. 
Using the theory, we showed that 
the fluorescence spectrum of the KNR shows characteristic 
features distinct from the two-level system in the strong-drive regime, 
where the KNR is driven strongly such that the Rabi frequency is comparable or larger than the Kerr nonlinearity. 
Those features include the appearance of additional subpeaks, where higher excited states of the KNR 
are involved, the decrease of the center-peak intensity as the drive power, and asymmetric intensity of the 
sideband peaks in the presence of dephasing. 
To gain insight on those features, we  calculated the steady-state population of the initial dressed state 
and its transition matrix element to the final dressed state and showed that 
their product gives a good estimate of the relative intensity of the corresponding peaks in the fluorescence spectrum. 

We experimentally measured the resonance fluorescence using a superconducting KNR in the 
strong-drive regime, where both the Kerr nonlinearity and the Rabi frequency are $\sim10$~MHz. 
We confirmed the decrease of the center-peak intensity. 
The overall spectra at different drive powers are well reproduced by our theory. 

Future study includes the similar experiment at non-zero flux bias point to investigate how the spectrum changes as 
a function of the flux bias, namely, the strength of the dephasing rate. It will help to understand the 
basic properties of the KNR, which is an interesting platform for the quantum information processing~\cite{Goto_2016b, Goto_2016, Puri_2017b, Puri_2017}. 

\begin{acknowledgements}
We thank S. Goto, Y. Urade and Y. Kubo for useful discussions. 
We also thank Y. Kitagawa for his assistance in the device fabrication and 
I. Takase for his contribution at the early stage of this work. 
The devices were fabricated in the Superconducting Quantum Circuit Fabrication Facility
(Qufab) in National Institute of Advanced Industrial Science and Technology (AIST). 
Part of this work was conducted at AIST Nano-Processing Facility supported by 
“Nanotechnology Platform Program” of the Ministry of Education, Culture, Sports, Science and Technology(MEXT), Japan. 
This paper is based on results obtained from a project, JPNP16007, commissioned by the New Energy and Industrial Technology
Development Organization (NEDO).
\end{acknowledgements}

\onecolumngrid
\appendix
\section{Calculation of fluorescence spectrum}~\label{App1}
Here we show the detailed calculation of the fluorescence spectrum in a system shown in Fig.~\ref{fig1} in the main text. 
To be self-contained, we also show the calculation of the fluorescence spectrum for the ideal two-level system, 
although it can be found in previous reports~\cite{Koshino_2012, Lu_2021}.

\subsection{Kerr nonlinear resonator}
We consider the Hamiltonian Eq.~(\ref{H_sys}) in the main text. 
From the Heisenberg equations of motion for $b_k$, we obtain 
\begin{equation}\label{Heq_b}
\frac{d}{dt}b_k(t) = -\mathrm{i} vk b_k(t) -\mathrm{i} \sqrt{\frac{v\kappa_{\rm ex}}{2 \pi}}\, a(t). 
\end{equation} 
By formally solving this differential equation, 
we have
\begin{equation}\label{sol_Heq_b}
b_k(t) = e^{- \mathrm{i} vkt} b_k(0) 
- \mathrm{i} \sqrt{\frac{v\kappa_{\rm ex}}{2 \pi}} \int_{0}^{t} \! dt' \, e^{- \mathrm{i} vk(t - t')} a(t'). 
\end{equation} 
We introduce the real-space representation of the waveguide field as
\begin{equation}
\widetilde{b}_r=\frac{1}{\sqrt{2\pi}}\int \! dk \, e^{\mathrm{i}kr} b_k.
\end{equation}
Note that $\bigl[ \widetilde{b}_r, \widetilde{b}^\dagger_{r'} \bigr] = \delta (r-r')$. 
In this representation, 
the waveguide field interacts with the resonator at $r=0$, and 
the $r<0$ ($r>0$) region corresponds to the incoming (outgoing) field.
From Eq.~(\ref{sol_Heq_b}), we have
\begin{eqnarray}~\nonumber
\widetilde{b}_r(t) 
&=& \frac{1}{\sqrt{2\pi}}\int \! dk \, e^{\mathrm{i}k(r-vt)} b_k(0)
- \mathrm{i} \frac{\sqrt{v\kappa_{\rm ex}}}{2 \pi} \int^t_0 \! dt' \int \! dk \, e^{\mathrm{i}k[r-v(t-t')]} a(t')\\~\nonumber
&=& \widetilde{b}_{r-vt}(0) 
- \mathrm{i} \sqrt{v\kappa_{\rm ex}} \int^t_0 \! dt' \, \delta[r-v(t-t')] a(t') \\ ~\nonumber
&=& \widetilde{b}_{r-vt}(0) 
- \mathrm{i} \sqrt{\frac{\kappa_{\rm ex}}{v}} \int^t_0 \! dt' \, \delta(t'-t+r/v) a(t') \\ 
&=& \widetilde{b}_{r-vt}(0) 
- \mathrm{i} \sqrt{\frac{\kappa_{\rm ex}}{v}}\, \theta(r)\theta(t-r/v)a(t-r/v) \label{br_op}, 
\end{eqnarray}
where $\theta(r)$ is the Heaviside step function. 
We define the input and output operators as 
\begin{eqnarray} \label{bin_op}
b_{\rm in}(t) &\equiv& \widetilde{b}_{-0}(t) 
= \widetilde{b}_{-vt}(0), \\ \label{bout_op}
b_{\rm out}(t) &\equiv& \widetilde{b}_{+0}(t) 
= b_{\rm in}(t) - \mathrm{i} \sqrt{\frac{\kappa_{\rm ex}}{v}}\, a(t). 
\end{eqnarray}
Using Eqs.~(\ref{br_op}) and (\ref{bin_op}), 
the field operator $\widetilde{b}_r(t)$ at the resonator position ($r=0$) is given by 
\begin{equation} \label{b0_op}
\widetilde{b}_0(t) = \frac{1}{\sqrt{2\pi}}\int \! dk \, b_k(t) = 
b_{\rm in}(t) - \frac{\mathrm{i}}{2}\sqrt{\frac{\kappa_{\rm ex}}{v}}\, a(t). 
\end{equation}
Similarly, 
from the Heisenberg equations of motion for $d_k$, we obtain 
\begin{equation}\label{Heq_d}
\frac{d}{dt}d_k(t) = -\mathrm{i} vk d_k(t) -\mathrm{i} \sqrt{\frac{v\gamma_{\rm p}}{\pi}}\, a^\dagger(t)a(t), 
\end{equation} 
and
\begin{equation}
d_k(t) = e^{- \mathrm{i} vkt} d_k(0) 
- \mathrm{i} \sqrt{\frac{v\gamma_{\rm p}}{\pi}} \int_{0}^{t} \! dt' \, e^{- \mathrm{i} vk(t - t')} a^\dagger(t')a(t'). 
\end{equation} 
The field operator $\widetilde{d}_r(t)$ at the resonator position ($r=0$) is given by 
\begin{equation} \label{d0_op}
\widetilde{d}_0(t) = \frac{1}{\sqrt{2\pi}}\int \! dk \, d_k(t) = 
d_{\rm in}(t) - \mathrm{i}\sqrt{\frac{\gamma_{\rm p}}{2v}}\, a^\dagger(t)a(t). 
\end{equation}

From the Heisenberg equations of motion for $a$, we then obtain 
\begin{eqnarray}~\nonumber
\frac{da}{dt} &=& 
\mathrm{i}[\mathcal{H}_{\rm sys}(t)/\hbar,a]
+ \mathrm{i} \sqrt{v\kappa_{\rm ex}}\, [a^\dagger,a] \widetilde{b}_0
+ \mathrm{i} \sqrt{v\kappa_{\rm in}}\, [a^\dagger,a] \widetilde{c}_0
+ \mathrm{i} \sqrt{2v\gamma_{\rm p}}\, \Bigl([a^\dagger a,a] \widetilde{d}_0 + \widetilde{d}^\dagger_0 [a^\dagger a,a] \Bigr) \\
&=& 
\mathrm{i}[\mathcal{H}_{\rm sys}(t)/\hbar,a]
- \mathrm{i}\sqrt{v\kappa_{\rm ex}}\, \widetilde{b}_0
- \mathrm{i}\sqrt{v\kappa_{\rm in}}\, \widetilde{c}_0
- \mathrm{i}\sqrt{2v\gamma_{\rm p}}\, (a \widetilde{d}_0 + \widetilde{d}^\dagger_0 a), \label{Heq_a}
\end{eqnarray}
where we used $[a^\dagger a, a]=-a$. 
Using Eq.~(\ref{b0_op}) and its counterparts for $\widetilde{c}_0$ and $\widetilde{d}_0$, 
Eq.~(\ref{Heq_a}) is rewritten as 
\begin{eqnarray}~\nonumber
\frac{da}{dt} &=& \label{Heq_a2_pre}
\mathrm{i}[\mathcal{H}_{\rm sys}(t)/\hbar,a]-\frac{\kappa_{\rm ex} + \kappa_{\rm in}}{2}a +\gamma_{\rm p} [a^\dagger a, a] 
- \mathrm{i}\sqrt{v\kappa_{\rm ex}}\, b_{\rm in}(t) 
- \mathrm{i}\sqrt{v\kappa_{\rm in}}\, c_{\rm in}(t) \\~\nonumber 
&& - \mathrm{i}\sqrt{2v\gamma_{\rm p}}\, [a d_{\rm in}(t) + d^\dagger_{\rm in}(t) a]\\~\nonumber
&=& -\Big(\mathrm{i}\Omega_0 + \mathrm{i} K a^\dagger a + \frac{\kappa}{2} \Bigr)a 
- \mathrm{i}\sqrt{v\kappa_{\rm ex}}\, b_{\rm in}(t) 
- \mathrm{i}\sqrt{v\kappa_{\rm in}}\, c_{\rm in}(t) \\
&& - \mathrm{i}\sqrt{2v\gamma_{\rm p}}\, [a d_{\rm in}(t) + d^\dagger_{\rm in}(t) a], \label{Heq_a2} 
\end{eqnarray}
where $\kappa=\kappa_{\rm ex} + \kappa_{\rm in} + 2\gamma_{\rm p}$ and we used $[ a^\dagger a^\dagger aa,a] = -2 a^\dagger aa$. 
We assume that at $t$=0, the system is in the vacuum and a drive field $\mathcal{F}(r/v-t)$ is injected from the external port, where
\begin{equation}~\label{mathcalF}
\mathcal{F}(r/v-t) = Fe^{\mathrm{i}\omega_{\rm d}(r/v-t)}/\sqrt{v}.
\end{equation}
%
The initial state vector is represented as 
\begin{equation}~\label{isv}
|\psi(0)\rangle = \mathcal{N} \exp\Bigl(\int_{-\infty}^0 \! dr \, \mathcal{F}(r/v)\widetilde{b}_r^\dagger(0)\Bigr) |{\rm v}\rangle, 
\end{equation}
where $|{\rm v}\rangle$ represents the ground state of the system, and $\mathcal{N}$ is a normalization constant. 
%
Note that $|\psi(0)\rangle$ is the eigenstate of $b_{\rm in}(t)$, namely, 
\begin{equation}~\label{bin_eigen}
b_{\rm in}(t)|\psi(0)\rangle = \widetilde{b}_{-vt}(0)|\psi(0)\rangle  = \mathcal{F}(-t)|\psi(0)\rangle. 
\end{equation}
Taking the expectation value of Eq.~(\ref{Heq_a2}) with the initial state of the system $|\psi(0)\rangle$, 
we obtain
\begin{eqnarray}~\nonumber
\frac{d\langle a \rangle}{dt} &=& 
-\Big(\mathrm{i}\Omega_0 + \frac{\kappa}{2} \Bigr) \langle a \rangle -\mathrm{i}K \langle a^\dagger aa \rangle
- \mathrm{i}\sqrt{v\kappa_{\rm ex}}\, \langle b_{\rm in}(t) \rangle 
- \mathrm{i}\sqrt{v\kappa_{\rm in}}\, \langle c_{\rm in}(t) \rangle \\~\nonumber
&& - \mathrm{i}\sqrt{2v\gamma_{\rm p}}\, [\langle a d_{\rm in}(t) \rangle +  \langle d^\dagger_{\rm in}(t) a \rangle] \\
&=& \label{eqm_a}
-\Big(\mathrm{i}\Omega_0 + \frac{\kappa}{2} \Bigr) \langle a \rangle -\mathrm{i}K \langle a^\dagger aa \rangle
- \mathrm{i}\sqrt{v\kappa_{\rm ex}}\, \langle b_{\rm in}(t) \rangle, 
\end{eqnarray}
where we used $\langle c_{\rm in}(t) \rangle=0$, $d_{\rm in}(t)|\psi(0) \rangle=0$
and $\langle \psi(0)|d_{\rm in}^\dagger(t)=0$. 
As seen from Eq.~(\ref{eqm_a}), we cannot construct 
simultaneous equations of motion in a closed form. 
Thus we consider the equation of motion for $A_{m,n} \equiv {a^\dagger}^m a^n$ and 
numerically solve simultaneous equations of motion for finite $m$ and $n$.  
By using relations such as $\bigl[a^\dagger, a^n\bigr] = -na^{n-1}$ and $\bigl[{a^\dagger}^2,a^n\bigr] = -n(n-1)a^{n-2}-2na^\dagger a^{n-1}$, 
the equation of motion for $A_{m,n}$ is given by. 
\begin{eqnarray} \nonumber
\frac{dA_{m,n}(t)}{dt}&=&
\Big[\mathrm{i}(m-n)\Omega_0 + \mathrm{i}(m-n)(m+n-1)\frac{K}{2} 
-(n+m)\frac{\kappa_{\rm ex}+\kappa_{\rm in}}{2} - (m-n)^2\gamma_{\rm p} \Bigr]A_{m,n}(t) \\ \nonumber
&& +\mathrm{i}K(m-n)A_{m+1,n+1}(t)\\ \nonumber
&& - \mathrm{i}\sqrt{v\kappa_{\rm ex}}\, [nA_{m,n-1}(t)b_{\rm in}(t)-
mb^\dagger_{\rm in}(t)A_{m-1,n}(t)] \\ \nonumber
&& - \mathrm{i}\sqrt{v\kappa_{\rm in}}\, [nA_{m,n-1}(t)c_{\rm in}(t)-
mc^\dagger_{\rm in}(t)A_{m-1,n}(t)] \\ \label{eom_Amn}
&& + \mathrm{i}\sqrt{2v\gamma_{\rm p}}\, (m-n) [A_{m,n}(t)d_{\rm in}(t)+
d^\dagger_{\rm in}(t)A_{m,n}(t)]. 
\end{eqnarray}
We take the expectation value of both sides of Eq.~(\ref{eom_Amn}) with $|\psi(0)\rangle$ and obtain 
\begin{multline}
\frac{d\langle A_{m,n}(t) \rangle}{dt}=
\epsilon_{m,n}\langle A_{m,n}(t) \rangle +\mathrm{i}K(m-n)\langle A_{m+1,n+1}(t) \rangle\\~\label{eqm_Ae}
 - \mathrm{i}\sqrt{\kappa_{\rm ex}}\, [n\langle A_{m,n-1}(t)\rangle F e^{-\mathrm{i}\omega_{\rm d}t}
-m\langle A_{m-1,n}(t)\rangle F^* e^{\mathrm{i}\omega_{\rm d}t}],  
\end{multline}
where we define
\begin{equation}
\epsilon_{m,n} \equiv\mathrm{i}(m-n)\Omega_0 + \mathrm{i}(m-n)(m+n-1)\frac{K}{2} 
-(m+n)\frac{\kappa_{\rm ex}+\kappa_{\rm in}}{2} - (m-n)^2\gamma_{\rm p}.
\end{equation}
Assuming a steady state, we set $\langle A_{m,n}(t)\rangle$ as 
\begin{equation}~\label{Amn}
\langle A_{m,n}(t)\rangle = \langle\alpha_{m,n}\rangle_{\rm s} e^{\mathrm{i}(m-n)\omega_{\rm d}t}
\end{equation}
and substitute it into Eq.~(\ref{eqm_Ae}), which leads to 
\begin{equation}~\label{ss_alpha}
0=\frac{d}{dt}\langle \alpha_{m,n}\rangle_{\rm s} = 
\epsilon_{m,n}' \langle \alpha_{m,n}\rangle_{\rm s} +\mathrm{i}K(m-n) \langle \alpha_{m+1,n+1}\rangle_{\rm s}
- \mathrm{i}\sqrt{\kappa_{\rm ex}}\, [n \langle \alpha_{m,n-1}\rangle_{\rm s} F
- m \langle \alpha_{m-1,n}\rangle_{\rm s} F^*],
\end{equation} 
where
\begin{equation}
\epsilon_{m,n}' \equiv 
\epsilon_{m,n} - \mathrm{i}(m-n)\omega_{\rm d} =
\mathrm{i}(m-n)(\Omega_0-\omega_{\rm d}) + \mathrm{i}(m-n)(m+n-1)\frac{K}{2} 
-(m+n)\frac{\kappa_{\rm ex}+\kappa_{\rm in}}{2} - (m-n)^2\gamma_{\rm p}.
\end{equation}
Therefore, $\langle \alpha_{m,n}\rangle_{\rm s}$ is obtained by 
numerically solving the above equations
for $m,n=0,1,\cdots, N_{\rm max}$, except for $(m,n)=(0,0)$, where we set $\langle \alpha_{0,0}\rangle_{\rm s}=1$. 

Next we consider the spectrum of the output field $b_{\rm out}(t)$. 
The spectrum is given by
\begin{equation}~\label{spd}
S(\omega) =
\Re \int_0^\infty \! \frac{d\tau}{\pi} \, e^{\mathrm{i}\omega\tau}
\langle b_{\rm out}^\dagger(t) b_{\rm out}(t+\tau)\rangle. 
\end{equation}
Using Eq.~(\ref{bout_op}), 
\begin{eqnarray}~\nonumber 
\langle b_{\rm out}^\dagger(t) b_{\rm out}(t+\tau)\rangle
&=& 
\Big\langle \big[ b_{\rm in}^\dagger(t) + \mathrm{i} \sqrt{\frac{\kappa_{\rm ex}}{v}}\, a^\dagger(t) \big]
\big[ b_{\rm in}(t+\tau) - \mathrm{i} \sqrt{\frac{\kappa_{\rm ex}}{v}}\, a(t+\tau) \bigr] \Big\rangle \\~\nonumber
&=&
\big\langle b_{\rm in}^\dagger(t)b_{\rm in}(t+\tau)\big\rangle + 
\mathrm{i} \sqrt{\frac{\kappa_{\rm ex}}{v}}\, \big\langle a^\dagger(t) b_{\rm in}(t+\tau) \big\rangle \\~\nonumber 
&&
-\mathrm{i} \sqrt{\frac{\kappa_{\rm ex}}{v}}\, \big\langle b_{\rm in}^\dagger(t) a(t+\tau) \big\rangle + 
\frac{\kappa_{\rm ex}}{v} \big\langle a^\dagger(t)a(t+\tau) \big\rangle \\~\nonumber 
&=&
\big\langle b_{\rm in}^\dagger(t) \big\rangle \big\langle b_{\rm in}(t+\tau)\big\rangle + 
\mathrm{i} \sqrt{\frac{\kappa_{\rm ex}}{v}}\, \big\langle a^\dagger(t) \big\rangle \big\langle b_{\rm in}(t+\tau) \big\rangle -
\mathrm{i} \sqrt{\frac{\kappa_{\rm ex}}{v}}\, \big\langle b_{\rm in}^\dagger(t) \big\rangle \big\langle a(t+\tau) \big\rangle \\~\nonumber 
&& 
+\frac{\kappa_{\rm ex}}{v} \big\langle a^\dagger(t) \big\rangle \big\langle a(t+\tau) \big\rangle +
\frac{\kappa_{\rm ex}}{v} \big\langle a^\dagger(t), a(t+\tau) \big\rangle \\ \label{bout_2pt}
&=& 
\langle b_{\rm out}^\dagger(t)\rangle\langle b_{\rm out}(t+\tau)\rangle + 
\frac{\kappa_{\rm ex}}{v} \big\langle a^\dagger(t), a(t+\tau) \big\rangle,  
\end{eqnarray}
where in the third equality, we used Eq.~(\ref{bin_eigen}), and 
$\langle A, B\rangle = \langle AB\rangle - \langle A\rangle\langle B\rangle$. 
From Eqs.~(\ref{spd}) and (\ref{bout_2pt}), 
\begin{eqnarray}
S(\omega) &\equiv& S_{\rm c}(\omega) + S_{\rm i}(\omega) \\ \label{Scoherent}
S_{\rm c}(\omega) &=& 
\Re \int_0^\infty \! \frac{d\tau}{\pi} \, e^{\mathrm{i}\omega\tau}
\langle b_{\rm out}^\dagger(t)\rangle
\langle b_{\rm out}(t+\tau)\rangle \\ \label{Sincoherent}
S_{\rm i}(\omega) &=&
\Re \int_0^\infty \! \frac{d\tau}{\pi} \, e^{\mathrm{i}\omega\tau}\frac{\kappa_{\rm ex}}{v}
\langle a^\dagger(t),a(t+\tau)\rangle, 
\end{eqnarray}
where $S_{\rm c}(\omega)$ and $S_{\rm i}(\omega)$ represent 
coherent and incoherent components, respectively. 

First, we consider the incoherent component and define the following quantity. 
\begin{equation}
B_{m,n}(t,\tau) \equiv \langle a^\dagger(t), A_{m,n}(t+\tau) \rangle. 
\end{equation}
The equation of motion for $B_{m,n}$ is given by 
\begin{multline}~\label{eom_B}
\frac{d}{d\tau}B_{m,n}(t,\tau) =
\epsilon_{m,n} B_{m,n}(t,\tau) +\mathrm{i}K(m-n) B_{m+1,n+1}(t,\tau) \\
- \mathrm{i}\sqrt{\kappa_{\rm ex}}\, [n B_{m,n-1}(t,\tau) Fe^{-\mathrm{i}\omega_{\rm d}(t+\tau)}-
m B_{m-1,n}(t,\tau) F^*e^{\mathrm{i}\omega_{\rm d}(t+\tau)}]. 
\end{multline}
Assuming a steady state, we set 
\begin{equation}
B_{m,n}(t,\tau) = e^{\mathrm{i}(m-n)\omega_{\rm d}\tau}e^{\mathrm{i}(m-n+1)\omega_{\rm d}t} 
\beta_{m,n}(\tau),
\end{equation}
and substitute it into Eq.~(\ref{eom_B}) to obtain
\begin{equation}~\label{eom_beta}
\frac{d}{d\tau}\beta_{m,n} =
\epsilon_{m,n}' \beta_{m,n} +\mathrm{i}K(m-n) \beta_{m+1,n+1}
 - \mathrm{i}\sqrt{\kappa_{\rm ex}}\, [n \beta_{m,n-1} F- m \beta_{m-1,n} F^*]. 
\end{equation}
In order to solve this equation, we introduce Laplace transformation of $\beta_{m,n}$, 
\begin{equation}
\overline{\beta_{m,n}}(s) \equiv \int_0^\infty \! d\tau \, e^{-s\tau}\beta_{m,n}(\tau).
\end{equation}
By Laplace transforming both sides of Eq.~(\ref{eom_beta}), 
\begin{equation}~\label{L_eom_beta}
\beta_{m,n}(0) =
(s-\epsilon_{m,n}')\, \overline{\beta_{m,n}}(s)
-\mathrm{i}K(m-n)\, \overline{\beta_{m+1,n+1}}(s)
 + \mathrm{i}\sqrt{\kappa_{\rm ex}}\, [n\, \overline{\beta_{m,n-1}}(s) F-
m\, \overline{\beta_{m-1,n}}(s)F^*]. 
\end{equation}
Here $\beta_{m,n}(0)$ can be calculated as 
\begin{eqnarray}~\nonumber
\beta_{m,n}(0)
&=&
B_{m,n}(t,0)e^{-\mathrm{i}(m-n+1)\omega_{\rm d}t} \\~\nonumber
&=& \langle a^\dagger(t), A_{m,n}(t) \rangle e^{-\mathrm{i}(m-n+1)\omega_{\rm d}t} \\~\nonumber
&=& \Bigl[ \langle A_{m+1,n}(t) \rangle -\langle A_{1,0}(t) \rangle\langle A_{m,n}(t) \rangle \Bigr]e^{-\mathrm{i}(m-n+1)\omega_{\rm d}t} \\
&=& \langle \alpha_{m+1,n} \rangle_{\rm s} - \langle \alpha_{1,0} \rangle_{\rm s}\langle \alpha_{m,n} \rangle_{\rm s}, 
\end{eqnarray}
where we used Eq.~(\ref{Amn}) in the last equality. Thus, in the steady state, Eq.~(\ref{L_eom_beta}) becomes 
\begin{multline}~\label{L_eom_beta_alpha}
\langle \alpha_{m+1,n} \rangle_{\rm s} - \langle \alpha_{1,0} \rangle_{\rm s} \langle \alpha_{m,n} \rangle_{\rm s}=
(s-\epsilon’_{m,n})\, \overline{\beta_{m,n}}(s)
-\mathrm{i}K(m-n)\, \overline{\beta_{m+1,n+1}}(s) \\
 + \mathrm{i}\sqrt{\kappa_{\rm ex}}\, [n\, \overline{\beta_{m,n-1}}(s) F- m\, \overline{\beta_{m-1,n}}(s)F^*]. 
\end{multline}
Using the solution of Eq.~(\ref{ss_alpha}), we can numerically solve 
Eq.~(\ref{L_eom_beta_alpha}) for $m,n=0,1,\cdots, N_{\rm max}$, except for $\overline{\beta_{0,0}}(s)=0$. 
From Eq.~(\ref{Sincoherent}), 
\begin{eqnarray}~\nonumber
S_{\rm i}(\omega)
&=&
\Re \int_0^\infty \! \frac{d\tau}{\pi} \, e^{\mathrm{i}\omega\tau}\frac{\kappa_{\rm ex}}{v}
\langle a^\dagger(t),a(t+\tau)\rangle \\~\nonumber
&=& \frac{\kappa_{\rm ex}}{\pi v}\Re \int_0^\infty \! d\tau \, e^{\mathrm{i}\omega\tau}B_{0,1}(t, \tau) \\
&=& \frac{\kappa_{\rm ex}}{\pi v}\Re \overline{\beta_{0,1}}\bigl(\mathrm{i}(\omega_{\rm d}-\omega)\bigr)~\label{Si_kpo}. 
\end{eqnarray}

Next we consider the coherent component. In the steady state, 
\begin{eqnarray}~\nonumber
\langle b_{\rm out}^\dagger(t)\rangle
\langle b_{\rm out}(t+\tau)\rangle
&=& \Big\langle b_{\rm in}^\dagger(t) + \mathrm{i} \sqrt{\frac{\kappa_{\rm ex}}{v}}\, a^\dagger(t) \Big\rangle
\Big\langle b_{\rm in}(t+\tau) - \mathrm{i} \sqrt{\frac{\kappa_{\rm ex}}{v}}\, a(t+\tau) \Big\rangle \\~\nonumber 
&=& \frac{e^{-\mathrm{i}\omega_{\rm d}\tau}}{v}
(F - \mathrm{i} \sqrt{\kappa_{\rm ex}}\, \langle \alpha_{0,1}\rangle_{\rm s})
(F^* + \mathrm{i} \sqrt{\kappa_{\rm ex}}\, \langle \alpha_{1,0}\rangle_{\rm s}). \\
&=& \frac{e^{-\mathrm{i}\omega_{\rm d}\tau}}{v} \Bigl|F - \mathrm{i} \sqrt{\kappa_{\rm ex}}\, \langle \alpha_{0,1}\rangle_{\rm s}\Bigr|^2.
\end{eqnarray}
From Eq.~(\ref{Scoherent}), 
\begin{eqnarray}~\nonumber
S_{\rm c}(\omega)
&=&
\Re \int_0^\infty \! \frac{d\tau}{\pi} \, e^{\mathrm{i}\omega\tau}
\langle b_{\rm out}^\dagger(t)\rangle
\langle b_{\rm out}(t+\tau)\rangle \\
&=& 
\frac{\delta(\omega-\omega_{\rm d})}{v}\Bigl|F - \mathrm{i} \sqrt{\kappa_{\rm ex}}\, \langle \alpha_{0,1}\rangle_{\rm s}\Bigr|^2. ~\label{Sc_KPO}
\end{eqnarray}

%
%

\subsection{Ideal two-level system}
In this section, we consider the case where the Kerr nonlinear resonator in Fig.~\ref{fig1} is replaced by an ideal two-level system~\cite{Koshino_2012, Lu_2021}.
Namely, 
\begin{equation}~\label{H_sys_tls}
\mathcal{H}_{\rm sys}(t)/\hbar = \Omega_0 \sigma^\dagger \sigma, 
\end{equation}
Similarly to the previous section, the equation of motion for $\sigma$ and $\sigma^\dagger \sigma$ is given by 
\begin{eqnarray}~\nonumber
\frac{d\sigma}{dt} &=&
-\Big(\mathrm{i}\Omega_0 + \frac{\kappa}{2} \Bigr)\sigma 
- \mathrm{i}\sqrt{v\kappa_{\rm ex}}\, (1 - 2\sigma^\dagger \sigma) b_{\rm in}(t) 
- \mathrm{i}\sqrt{v\kappa_{\rm in}}\, (1 - 2\sigma^\dagger \sigma) c_{\rm in}(t)  \\ \label{eqm_s}
&&
- \mathrm{i}\sqrt{2v\gamma_{\rm p}}\, \bigl[ \sigma d_{\rm in}(t) + d^\dagger_{\rm in}(t) \sigma \bigr], \\~\nonumber
\frac{d}{dt}\sigma^\dagger \sigma &=&
- (\kappa_{\rm ex} + \kappa_{\rm in})\sigma^\dagger \sigma 
+ \mathrm{i}\sqrt{v\kappa_{\rm ex}}\, \bigl[ b_{\rm in}^\dagger(t)\sigma - \sigma^\dagger b_{\rm in}(t)\bigr] \\
&&
+ \mathrm{i}\sqrt{v\kappa_{\rm in}}\, \bigl[ c_{\rm in}^\dagger(t)\sigma - \sigma^\dagger c_{\rm in}(t)\bigr], 
\end{eqnarray}
where we used $[ \sigma, \sigma^\dagger] = 1 - \sigma^\dagger \sigma$ and $[\sigma^\dagger \sigma, \sigma] = -\sigma$. 
Taking the expectation value of $|\psi(0)\rangle$ on both sides, we get 
\begin{eqnarray}~\label{eqm_s_exp}
\frac{d\langle \sigma \rangle}{dt} &=&
-\Big(\mathrm{i}\Omega_0 + \frac{\kappa}{2} \Bigr)\langle \sigma \rangle 
- \mathrm{i}\sqrt{v\kappa_{\rm ex}}\, (1 - 2\langle \sigma^\dagger \sigma \rangle) \mathcal{F}(-t), \\~\label{eqm_sds_exp}
\frac{d}{dt}\langle \sigma^\dagger \sigma \rangle &=&
- (\kappa_{\rm ex} + \kappa_{\rm in})\langle \sigma^\dagger \sigma \rangle 
+ \mathrm{i}\sqrt{v\kappa_{\rm ex}}\, \bigl[ \mathcal{F}^*(-t)\langle\sigma\rangle - \mathcal{F}(-t) \langle\sigma\rangle^* \bigr]. 
\end{eqnarray}
Note that in contrast to Eq.~(\ref{eqm_a}), Eqs.~(\ref{eqm_s_exp}) and (\ref{eqm_sds_exp}) form closed simultaneous equations. 
To obtain the steady-state solutions, we set 
\begin{eqnarray}~\label{s_ss}
\langle\sigma(t)\rangle &=& \langle\sigma\rangle_{\rm s} e^{-\mathrm{i} \omega_{\rm d}t}, \\~\label{sds_ss}
\langle \sigma^\dagger(t) \sigma(t) \rangle &=& \langle \sigma^\dagger \sigma \rangle_{\rm s}.  
\end{eqnarray}
By substituting Eqs.~(\ref{mathcalF}), (\ref{s_ss}), and (\ref{sds_ss}) into Eqs.~(\ref{eqm_s_exp}) and (\ref{eqm_sds_exp}), 
we obtain 
\begin{eqnarray}~\label{sig_s}
\langle\sigma\rangle_{\rm s} &=& 
\frac{-\mathrm{i}\sqrt{\kappa_{\rm ex}}\, \bigl[\kappa/2 + \mathrm{i}(\omega_{\rm d} - \Omega_0)\bigr] F}
{2\overline{r}\kappa_{\rm ex}|F|^2 + |\kappa/2 + \mathrm{i}(\omega_{\rm d} - \Omega_0)|^2} \\~\label{sigd_sig_s}
\langle \sigma^\dagger \sigma \rangle_{\rm s} &=& 
\frac{\overline{r}\kappa_{\rm ex} |F|^2}
{2\overline{r}\kappa_{\rm ex}|F|^2 + |\kappa/2 + \mathrm{i}(\omega_{\rm d} - \Omega_0)|^2}, 
\end{eqnarray}
where $\overline{r}=\kappa/(\kappa_{\rm ex} + \kappa_{\rm in})$.

As seen from Eq.~(\ref{Sincoherent}), we need $\langle \sigma^\dagger(t), \sigma(t + \tau) \rangle$ to calculate the fluorescence, 
so that we define
\begin{equation}
B_1(t, \tau) \equiv \langle \sigma^\dagger(t), \sigma(t + \tau) \rangle
\end{equation}
and consider its equation of motion. 
We find that by additionally defining
\begin{eqnarray}
B_2(t, \tau) &\equiv& \langle \sigma^\dagger(t), \sigma^\dagger(t + \tau) \rangle \\
B_3(t, \tau) &\equiv& \langle \sigma^\dagger(t), \sigma^\dagger(t + \tau)\sigma(t + \tau) \rangle, 
\end{eqnarray} 
we obtain the simultaneous equations of motion in closed form as follows: 
\begin{eqnarray}~\nonumber
\frac{\partial}{\partial \tau} B_1(t,\tau) &=& 
\Bigl\langle\sigma^\dagger(t)\frac{\partial}{\partial\tau}\sigma(t+\tau)\Bigr\rangle - 
\bigl\langle\sigma^\dagger(t)\bigr\rangle\Bigl\langle\frac{\partial}{\partial\tau}\sigma(t+\tau)\Bigr\rangle \\ \nonumber
&=& 
\Bigl\langle\sigma^\dagger(t) \Bigl[-\Big(\mathrm{i}\Omega_0 + \frac{\kappa}{2} \Bigr)\sigma(t + \tau) 
- \mathrm{i}\sqrt{v\kappa_{\rm ex}}\, \bigl[1 - 2\sigma^\dagger(t+\tau) \sigma(t+\tau)\bigr] b_{\rm in}(t+\tau)\Bigr] \Bigr\rangle \\ \nonumber \label{B1_deriv}
&& -\bigl\langle\sigma^\dagger(t)\bigr\rangle \Bigl\langle-\Big(\mathrm{i}\Omega_0 + \frac{\kappa}{2} \Bigr)\sigma(t + \tau) 
- \mathrm{i}\sqrt{v\kappa_{\rm ex}}\, \bigl[1 - 2\sigma^\dagger(t+\tau) \sigma(t+\tau)\bigr] b_{\rm in}(t+\tau) \Bigr\rangle \\ \label{B1_deriv2}
&=& -\Big(\mathrm{i}\Omega_0 + \frac{\kappa}{2} \Bigr)B_1(t,\tau) + 2\mathrm{i}\sqrt{v\kappa_{\rm ex}}\, \mathcal{F}[-(t+\tau)]B_3(t,\tau) \\  \label{B2_deriv}
\frac{\partial}{\partial \tau} B_2(t,\tau) &=& 
\Big(\mathrm{i}\Omega_0 - \frac{\kappa}{2} \Bigr)B_2(t,\tau) - 2\mathrm{i}\sqrt{v\kappa_{\rm ex}}\, \mathcal{F}^*[-(t+\tau)]B_3(t,\tau) \\ \nonumber \label{B3_deriv}
\frac{\partial}{\partial \tau} B_3(t,\tau) &=& 
\mathrm{i}\sqrt{v\kappa_{\rm ex}}\, \mathcal{F}^*[-(t+\tau)]B_1(t,\tau) - \mathrm{i}\sqrt{v\kappa_{\rm ex}}\, \mathcal{F}[-(t+\tau)]B_2(t,\tau) \\
&& -(\kappa_{\rm ex}+\kappa_{\rm in}) B_3(t,\tau),
\end{eqnarray}
where in Eq.~(\ref{B1_deriv}), we used Eq.~(\ref{eqm_s}) and omitted terms containing $c_{\rm in}$ and $d_{\rm in}$, which will disappear in Eq.~(\ref{B1_deriv2}) after all. 
Assuming a steady state, we set 
\begin{eqnarray}
B_1(t, \tau) &=& e^{-\mathrm{i}\omega_{\rm d}\tau} \beta_1(\tau) \\
B_2(t, \tau) &=& e^{\mathrm{i}\omega_{\rm d}(2t+\tau)} \beta_2(\tau) \\
B_3(t, \tau) &=& e^{\mathrm{i}\omega_{\rm d}t} \beta_3(\tau), \\
\end{eqnarray}
and substitute them into the above equations with Eq.~(\ref{mathcalF}) and obtain 

\begin{equation}~\label{eom_beta_tls}
\frac{d}{d\tau} 
\begin{bmatrix} 
\beta_1(\tau) \\
\beta_2(\tau) \\
\beta_3(\tau) \\
\end{bmatrix}
=
\begin{bmatrix}
\mathrm{i}(\omega_{\rm d} - \Omega_0)-\kappa/2 & 0 & 2\mathrm{i}\sqrt{\kappa_{\rm ex}}\, F \\
0 & -\mathrm{i}(\omega_{\rm d} - \Omega_0)-\kappa/2 & -2\mathrm{i}\sqrt{\kappa_{\rm ex}}\, F^* \\
\mathrm{i}\sqrt{\kappa_{\rm ex}}\, F^* & -\mathrm{i}\sqrt{\kappa_{\rm ex}}\, F & -(\kappa_{\rm ex}+\kappa_{\rm in}) \\
\end{bmatrix} 
\begin{bmatrix} 
\beta_1(\tau) \\
\beta_2(\tau) \\
\beta_3(\tau) \\
\end{bmatrix}. 
\end{equation}
In order to solve this equation, we introduce Laplace transformation of $\beta_i(\tau)$ ($i$=1, 2 and 3), 
\begin{equation}
\overline{\beta_i}(s) \equiv \int_0^\infty \! d\tau \, e^{-s\tau}\beta_i(\tau).
\end{equation}
By Laplace transforming both sides of Eq.~(\ref{eom_beta_tls}), we obtain
\begin{equation}
\begin{bmatrix} 
\beta_1(0) \\
\beta_2(0) \\
\beta_3(0) \\
\end{bmatrix}
=
\begin{bmatrix}
s-\mathrm{i}(\omega_{\rm d} - \Omega_0) + \kappa/2 & 0 & -2\mathrm{i}\sqrt{\kappa_{\rm ex}}\, F \\
0 & s + \mathrm{i}(\omega_{\rm d} - \Omega_0)+\kappa/2 & 2\mathrm{i}\sqrt{\kappa_{\rm ex}}\, F^* \\
-\mathrm{i}\sqrt{\kappa_{\rm ex}}\, F^* & \mathrm{i}\sqrt{\kappa_{\rm ex}}\, F & s + \kappa_{\rm ex}+\kappa_{\rm in} \\
\end{bmatrix} 
\begin{bmatrix} 
\overline{\beta_1}(s) \\
\overline{\beta_2}(s) \\
\overline{\beta_3}(s). \\
\end{bmatrix}
\end{equation}
By noting that 
\begin{eqnarray}
\beta_1(0) &=& \langle \sigma^\dagger \sigma \rangle_{\rm s} - |\langle \sigma \rangle_{\rm s}|^2 \\
\beta_2(0) &=& -(\langle \sigma \rangle_{\rm s}^*)^2 \\
\beta_3(0) &=& -\langle \sigma \rangle_{\rm s}^* \langle \sigma^\dagger \sigma \rangle_{\rm s},  
\end{eqnarray}
where we used $\langle \sigma^\dagger(t) \sigma^\dagger(t) \rangle = 0$, we obtain 
\begin{equation}~\label{L_beta_tls}
\begin{bmatrix} 
\overline{\beta_1}(s) \\
\overline{\beta_2}(s) \\
\overline{\beta_3}(s) \\
\end{bmatrix}
=
\begin{bmatrix}
s-\mathrm{i}(\omega_{\rm d} - \Omega_0) + \kappa/2 & 0 & -2\mathrm{i}\sqrt{\kappa_{\rm ex}}\, F \\
0 & s + \mathrm{i}(\omega_{\rm d} - \Omega_0)+\kappa/2 & 2\mathrm{i}\sqrt{\kappa_{\rm ex}}\, F^* \\
-\mathrm{i}\sqrt{\kappa_{\rm ex}}\, F^* & \mathrm{i}\sqrt{\kappa_{\rm ex}}\, F & s + \kappa_{\rm ex}+\kappa_{\rm in} \\
\end{bmatrix}^{-1}
\begin{bmatrix} 
\langle \sigma^\dagger \sigma \rangle_{\rm s} - |\langle \sigma \rangle_{\rm s}|^2 \\
-(\langle \sigma \rangle_{\rm s}^*)^2 \\
-\langle \sigma \rangle_{\rm s}^* \langle \sigma^\dagger \sigma \rangle_{\rm s} \\
\end{bmatrix}.
\end{equation}

From Eq.~(\ref{Sincoherent}), 
\begin{eqnarray}~\nonumber
S_{\rm i}(\omega)
&=&
\Re \int_0^\infty \! \frac{d\tau}{\pi} \, e^{\mathrm{i}\omega\tau}\frac{\kappa_{\rm ex}}{v}
\langle \sigma^\dagger(t), \sigma(t+\tau)\rangle \\ \label{Si_tls}
&=& \frac{\kappa_{\rm ex}}{\pi v} \Re \overline{\beta_1}\bigl(\mathrm{i}(\omega_{\rm d} - \omega) \bigr).
\end{eqnarray}
By combining Eqs.~(\ref{L_beta_tls}) and (\ref{Si_tls}), we can numerically calculate $S_{\rm i}(\omega)$. 
As for $S_{\rm c}(\omega)$, we can calculate it similarly to Eq.~(\ref{Sc_KPO}) as 
\begin{equation}
S_{\rm c}(\omega) = \frac{\delta(\omega-\omega_{\rm d})}{v}\Bigl|F - \mathrm{i} \sqrt{\kappa_{\rm ex}}\, \langle \sigma\rangle_{\rm s}\Bigr|^2,  
\end{equation}
where $\langle \sigma\rangle_{\rm s}$ is given in Eq.~(\ref{sig_s}). 

\section{Calculation of the dressed-state populations}~\label{App2}
In this Appendix, we derive the relation between the density matrix in eigenstate basis and 
$\langle\alpha_{m,n}\rangle_{\rm s}$ in Eq.~(\ref{alpha_mn_s}) to calculate the population shown in Figs.~\ref{fig4}(b) and \ref{fig4}(c).
We denote the $m$th Fock state as $|m\rangle$ and the $i$th eigenstate of the Hamiltonian Eq.~(\ref{H_rot}) as $|\tilde{i}\rangle$, 
which are related by a unitary matrix 
\begin{equation}
U = \sum_{m,i}U_{mi}|m\rangle\langle\tilde{i}|.
\end{equation}
Using 
\begin{equation}
a^n = \sum_{k \ge 0}\sqrt{\frac{(k+n)!}{k!}}\, |k\rangle\langle k+n|,
\end{equation}
$\langle\alpha_{m,n}\rangle_{\rm s}$ is calculated as 
\begin{eqnarray}~\nonumber
\langle {a^\dagger}^m a^n \rangle_{\rm s} &=&  \sum_{k \ge 0}\frac{\sqrt{(k+m)!(k+n)!}}{k!}\, \rho_{n+k, m+k} \\
&=&  \sum_{k \ge 0}\frac{\sqrt{(k+m)!(k+n)!}}{k!}\, (U\tilde{\rho}U^\dagger)_{n+k, m+k},~\label{alpha_s} 
\end{eqnarray}
where $\rho$ and $\tilde{\rho}$ represent the density matrices in the Fock-state and dressed-state bases, respectively.
Introducing a diagonal and shifted-diagonal matrices of $D_{nm}=\sqrt{n!}\, \delta_{nm}$ and $A_{nm}=\delta_{n+1,m}$, 
Eq.~(\ref{alpha_s}) further leads
\begin{eqnarray}~\nonumber
\langle\alpha_{m,n}\rangle_{\rm s} &=& \frac{1}{k!}\sum_{k \ge 0}(DU\tilde{\rho}U^\dagger D)_{n+k, m+k} \\~\nonumber
&=&  \frac{1}{k!}\sum_{k \ge 0}(A^kDU\tilde{\rho}U^\dagger D{A^\dagger}^k)_{nm} \\
&=& (e^\mathcal{A}\cdot DU\tilde{\rho}U^\dagger D)_{nm},~\label{alpha_s_2} 
\end{eqnarray}
where $\mathcal{A}$ represents a linear mapping represented by $\mathcal{A}\cdot X \rightarrow AXA^\dagger$ for a matrix $X$.
From Eq.~(\ref{alpha_s_2}), 
\begin{eqnarray}~\nonumber
\tilde{\rho}_{ij} &=& [U^\dagger D^{-1}(e^{-\mathcal{A}}\cdot \alpha)D^{-1}U]_{ij} \\
&=& \sum_{m,n}\sum_{k \ge 0}\frac{(-1)^k}{k!\sqrt{n!m!}}U^\dagger_{im}\alpha_{n+k, m+k}U_{nj},~\label{alpha_s_3} 
\end{eqnarray}
where we defined a matrix $\alpha$ with elements given by $\alpha_{mn}=\langle {a^\dagger}^m a^n \rangle_{\rm s}$.
From Eq.~(\ref{alpha_s_3}), we obtain the population of the eigenstate as 
\begin{equation}
P_i=\tilde{\rho}_{ii}=\sum_{m,n}\sum_{k \ge 0}\frac{(-1)^k}{k!\sqrt{n!m!}}U^\dagger_{im}\alpha_{n+k, m+k}U_{ni}. 
\end{equation}

\begin{figure}[t]
\includegraphics[width=0.9\columnwidth,clip]{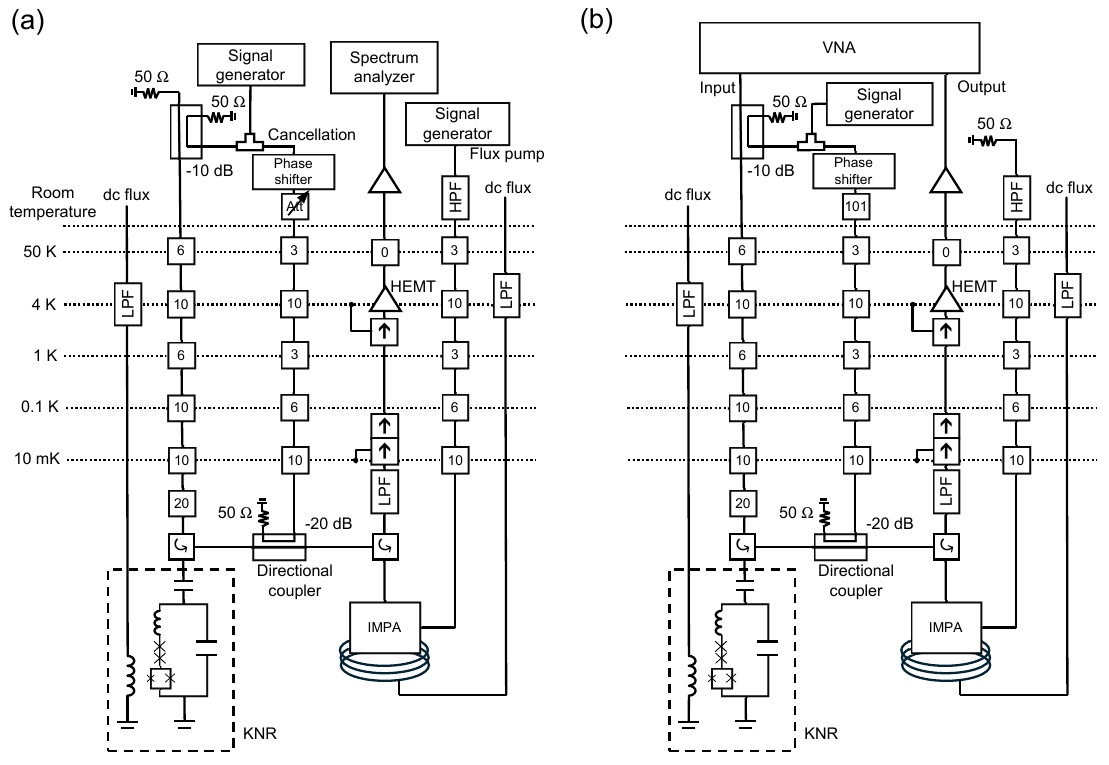}
\caption{~\label{fig9}
Experimental setup for the measurement of (a) fluorescence (Fig.~\ref{fig8}) and 
(b) reflection coefficient with and without drive field (Figs.~\ref{fig6} and \ref{fig7}).
}
\end{figure}

\section{Detailed experimental setup}~\label{App3}
Figure~\ref{fig9} shows the detailed measurement setup used in the present study. 
For the fluorescence measurement (Fig.~\ref{fig8}), 
we used the setup shown in Fig.~\ref{fig9}(a), 
where a spectrum analyzer was used as the detector. 
For the reflection-coefficient measurements with and without a drive field (Figs.~\ref{fig6} and \ref{fig7}), 
we used the setup shown in Fig.~\ref{fig9}(b), where a vector network analyzer (VNA) was used. 

In the fluorescence measurement, we used a cancellation field in order to suppress the reflected drive field 
by the destructive interference and avoid saturating IMPA. To adjust the amplitude and phase of the cancellation field, 
we used a phase shifter and a variable attenuator, which are put in one of the two paths from the output of the signal generator 
split by the power divider. 
After the adjustment, we suppressed the power below $\sim-160$~dBm at the input of IMPA.

In the reflection-coefficient measurements  (Figs.~\ref{fig6} and \ref{fig7}), 
we did not use IMPA. We turned off the pump and set the flux bias far detuned from the resonance frequency of KNR. 
Because the cancellation field is not necessary, we set the maximum attenuation of the variable attenuator (101~dB)
and effectively turned it off. 

\begin{figure}[hh]
\includegraphics[width=0.9\columnwidth,clip]{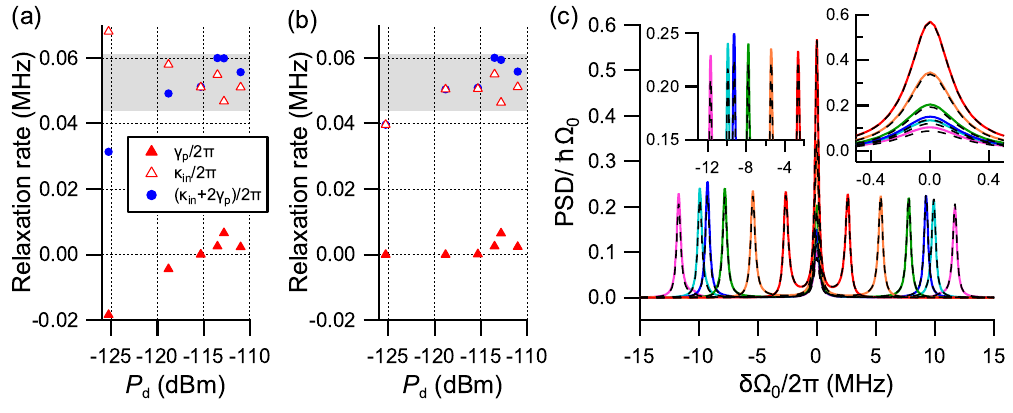}
\caption{~\label{fig10}
Relaxation rates obtained by fitting the fluorescence spectra using (a) the fmin function and (b) Optuna. 
The gray region represents the two standard deviations of the mean of $\kappa_{\rm in}^*$ 
determined from the reflection-coefficient measurements. 
(c) Comparison of the spectra with and without the dephasing. 
Solid curves represent the same data as in Fig.~\ref{fig8}(c), where $\gamma_{\rm p}/2\pi=0.0065$~MHz 
and $\kappa_{\rm in}/2\pi=0.046$~MHz, while the dashed curves represent those  
with $\gamma_{\rm p}$ and $\kappa_{\rm in}$ changed to zero and $2\pi \times 0.059$~MHz, respectively. 
The left upper inset shows the magnification of lower sideband peaks. The right upper inset shows the magnification of 
center peaks.  
}
\end{figure}
\section{Numerical fitting of the fluorescence data}~\label{App4}
In Fig.~\ref{fig8}(c) in the main text, we showed the numerical fitting of the fluorescence data with $P_{\rm d}=-112.8$~dBm 
using a python package Optuna~\cite{Akiba_2019}. 
We tried fitting the data with other $P_{\rm d}$'s with the same parameter setting. 
We also tried fitting using the fmin function in SciPy library with the initial guess of $A=1.0$, $\kappa_{\rm in}/2\pi=0.05$~MHz, 
and $\gamma_{\rm p}/2\pi=0.0025$~MHz. 
The obtained relaxation rates using the fmin function and Optuna 
as a function of $P_{\rm d}$ are plotted in Figs.~\ref{fig10}(a) and (b), respectively. 
For $P_{\rm d} < -115$~dBm, we could not obtain reasonable value of $\gamma_{\rm p}$ in either method. 
They are either too small 
or even negative in the case of the fmin function, where no restriction in the parameter range was set. 
We suspect that this is because of the lower sensitivity of the spectrum to $\gamma_{\rm p}$ at lower $P_{\rm d}$'s. 
Figure~\ref{fig10}(c) shows the comparison of the spectrum with and without the dephasing. 
Solid curves represent the same spectra as shown in Fig.~\ref{fig8}(c), where $\gamma_{\rm p}/2\pi=0.0065$~MHz 
and $\kappa_{\rm in}/2\pi=0.046$~MHz, while the dashed curves represent those 
with $\gamma_{\rm p}/2\pi=0$ and $\kappa_{\rm in}/2\pi=0.059$~MHz. 
As seen in the figure, their difference becomes smaller as $P_{\rm d}$ is decreased, 
which makes it more difficult to reliably extract $\gamma_{\rm p}$ by fitting the theoretical curve to the experimental data. 
For $P_{\rm d} > -115$~dBm, both fmin and Optuna give almost the same $\gamma_{\rm p}$ and $\kappa_{\rm in}$ 
for each $P_{\rm d}$. 
The calculated $\kappa_{\rm in}^*$s fall within two standard deviations~(indicated by the gray region in the figure) 
of the mean of $\kappa_{\rm in}^*$ determined from the reflection-coefficient measurements. 

\providecommand{\noopsort}[1]{}\providecommand{\singleletter}[1]{#1}%

\end{document}